\documentclass[onecolumn,amsmath,amssymb,nofootinbib,12pt]{article}
\pdfoutput=1
\usepackage{jheppub}
\usepackage{graphicx}% Include figure files
\usepackage{dcolumn}% Align table columns on decimal point
\usepackage{bm,psfrag}% bold math

\usepackage{subfigure}
\usepackage{float}
\usepackage{tensor}

\newcommand{\ben}{\begin{equation}}
\newcommand{\een}{\end{equation}}
\newcommand{\be}{\begin{equation}}
\newcommand{\ee}{\end{equation}}
\newcommand{\bea}{\begin{eqnarray}}
\newcommand{\eea}{\end{eqnarray}}
\newcommand{\ba}{\begin{eqnarray}}
\newcommand{\ea}{\end{eqnarray}}

\newcommand{\beq}{\begin{equation}}
\newcommand{\eeq}{\end{equation}}
\newcommand{\beqa}{\begin{eqnarray}}
\newcommand{\eeqa}{\end{eqnarray}}
\newcommand{\beqar}{\begin{eqnarray*}}
\newcommand{\eeqar}{\end{eqnarray*}}

\newcommand{\reef}[1]{(\ref{#1})}

\newcommand{\eg}{{\it e.g.,}\ }
\newcommand{\ie}{{\it i.e.,}\ }
\newcommand{\labell}[1]{\label{#1}} %{\mt{#1}\label{#1}} %

\newcommand{\cO}{{\cal O}}

\def\t6 {T_\mt{D6}}

\newcommand{\mt}[1]{\textrm{\tiny #1}}

\newcommand{\m}{v}

\newcommand{\vk}{{\vec{k}}}
\newcommand{\vx}{{\vec{x}}}

\def\cale         {{\cal E}}

\def\calo         {{\cal O}}

\def\ee           {{\rm e}}

\def\sqr#1#2{{\vcenter{\vbox{\hrule height.#2pt
 \hbox{\vrule width.#2pt height#1pt \kern#1pt
 \vrule width.#2pt}\hrule height.#2pt}}}}

\def\ee{\cale}

\def\aa1{\phi}
\def\cc1{\psi}

\def\k{\kappa}

\def\vev#1{\langle #1 \rangle}

\def\k{\kappa}

\def\nnn{\nonumber}

\def\tKZ{t_\mt{KZ}}
\def\tkz{t_\mt{KZ}}
\def\EKZ{E_\mt{KZ}}
\def\ekz{E_\mt{KZ}}

\newcommand{\dt}{\delta t}

\begin{document}

\preprint{arXiv:1602.08547 [hep-th]}

%\subheader{NOTES}

\title{Quantum Quenches in Free Field Theory: Universal Scaling at Any Rate}

\author{Sumit R. Das,$^{1}$ Dami\'an A. Galante$^{2,3}$ and Robert C. Myers$^3$}
\affiliation{$^1$\,Department of Physics and Astronomy, University of Kentucky,\\ 
\vphantom{k}\ \ Lexington, KY 40506, USA}
\affiliation{$^2$\,Department of Applied Mathematics, University of Western Ontario,\\ 
\vphantom{k}\ \ London, ON N6A 5B7, Canada}
\affiliation{$^3$\,Perimeter Institute for Theoretical Physics, Waterloo, ON N2L 2Y5, Canada}

\emailAdd{das@pa.uky.edu}
\emailAdd{dgalante@perimeterinstitute.ca}
\emailAdd{rmyers@perimeterinstitute.ca}

\date{\today}

\abstract{Quantum quenches display universal scaling in several regimes. For quenches which start from a gapped phase and cross a critical point, with a rate {\em slow} compared to the initial gap, many systems obey Kibble-Zurek scaling. More recently, a different scaling behaviour has been shown to occur when the quench rate is {\em fast} compared to all other physical scales, but still slow compared to the UV cutoff. We investigate the passage from fast to slow quenches in scalar and fermionic free field theories with time dependent masses for which the dynamics can be solved exactly for all quench rates. We find that renormalized one point functions smoothly cross over between the regimes.}

\date{\today}

\maketitle

%\tableofcontents

\section{Introduction}

Universal scaling behaviour is known to occur in quantum quench
processes which involve critical points. The best known example is
Kibble-Zurek (KZ) scaling \cite{kibble,zurek}, which has received considerable attention in the past several years \cite{more, gritsev}, including various holographic studies
\cite{holo-kz,holo-kz2}.
Consider, for example, a system with a
time dependent coupling $g(t)$ which is initially in a gapped
phase and whose subsequent time evolution takes it across a critical
point $g_c$ where the gap vanishes. Further, if the rate at which the coupling varies is {\em slow} compared to the initial gap, then the early time
evolution is essentially adiabatic. So if the system starts off in
the ground state of the initial Hamiltonian, then it continues to remain largely in the
instantaneous ground state. However as $g(t)$ approaches $g_c$, the instantaneous gap is approaching zero and so adiabaticity must break down and the system is excited. Consider the simple power-law protocol
%\footnote{In addition, in section \ref{ECP}, we will also analyze Kibble-Zurek behaviour in protocols that approach the critical point exponentially (instead of having a power law behaviour), \ie $g(t)-g_c \sim g_0\exp(-t/\dt)$.} 
\ben
g(t) - g_c \sim g_0\   \left( {t}/{\dt} \right)^{r} \,.
\label{0-1}
\een
The original arguments of
Kibble and Zurek \cite{kibble,zurek} (which were made for thermal transitions) are readily adapted to argue that immediately after the quench, \ie after entering the non-adiabatic regime, the expectation value of an operator $\cO_\Delta$ of dimension $\Delta$ will exhibit  universal scaling of the form (\eg see ref \cite{more})
\ben
\vev{\cO_\Delta} \sim \big( g_0  / \dt^r \big)^{\frac{\nu \Delta}{%z 
r \nu + 1}} \,,
\label{0-2}
\een
where %$z$ is the dynamical critical exponent and 
$\nu$ is the correlation length exponent.\footnote{That is, the exponent determining the instantaneous gap --- see eq.~\reef{1-2}. Throughout the paper, we will assume that the critical theory is relativistic, \ie the dynamical critical exponent is $z=1$.} In fact, in the vicinity of the critical point, the time dependence of $\vev{\cO_\Delta}$ is conjectured to be determined by a simple {\em scaling function} \cite{scfn,qcritkz} --- see eq.~\reef{t_over_tkz}. 
%\ben
%\vev{\cO_\Delta (t) }_{ren} \simeq \big( g_0  / \dt^r \big)^{\frac{\nu \Delta}{%z 
%r \nu + 1}} \,F(t~\left( \frac{g_0}{\dt^{\,r}} \right)^{\frac{\nu}{r \nu +1}} ) \,.
%\label{t_over_tkz-2}
%\een

An important aspect of the original Kibble-Zurek argument \cite{kibble,zurek} is a universal prediction for the scaling of density of defects for dynamics across an order-disorder transition. In this work, we will not address this issue. Rather we will be concerned with the evolution of the one point function of the quenched operator and the time in which this scaling will appear.\footnote{However, our results for mass quenches of free fermion theory, when applied to 1+1 dimensional Majorana fermions, can be adapted to a calculation of the kink density of the $1+1$ dimensional Ising model. This will be discussed in a future communication \cite{ddgms}.} Thus this paper deals with ``Kibble-Zurek scaling" rather than the ``Kibble-Zurek mechanism".

Recently a new scaling behaviour has also been found for {\em fast}
quenches. This scaling was first discovered in holographic studies
\cite{numer,fastQ}, but later shown to be a completely general result in any
quantum field theory whose UV limit is a conformal field theory \cite{dgm1,dgm2,dgm3}. Consider a quantum field
theory described by the action
\ben
S = S_{CFT} + \int dt ~\lambda (t) \int d^{d-1}x~\cO_\Delta (\vx,t) \,,
\label{0-3}
\een
where $\cO_\Delta$ is a relevant operator in the UV fixed point
theory with conformal dimension $\Delta$. The coupling $\lambda (t)$ 
starts from some constant value
$\lambda_1$, varies as a function of time over some time scale $\dt$
over a range of the order of $\delta \lambda$, 
and settles down to some other constant $\lambda_2$. When $\dt$ is
small compared to {\em all} other {\em physical} length scales 
in the problem, but {\em slow} compared
to the scale of the UV cutoff, \ie
\ben
 \Lambda_\mt{UV}^{-1}\ll \dt \ll (\lambda_1)^{1/(\Delta-d)}\,, (\lambda_2)^{1/(\Delta-d)}\,,  
(\delta \lambda)^{1/(\Delta-d)} \,,
\label{0-5}
\een
response of various {\em renormalized} quantities exhibits scaling. For example, during the quench process, the renormalized
expectation value $\vev{\cO_\Delta}_{ren}$ behaves as
\ben
\vev{\cO_\Delta}_{ren} \sim  \frac{\delta\lambda}{\dt^{2\Delta-d}} \,.
\label{0-4}
\een
This result matches the linear response theory result \cite{dgm2} and as a consequence of the diffeomorphism Ward identity, there is a similar scaling law for the energy density.\footnote{Further, since the energy is a conserved quantity, this scaling persists for all times after the quench.}
As discussed below, this early time scaling for smooth fast quenches can be
examined in great detail for free field theories \cite{dgm1,dgm2,dgm3}. In these cases, there are also additional scaling laws for higher spin conserved charges.

The aim of this paper is to investigate the transition from scaling in
fast quenches to Kibble-Zurek scaling as one changes the quench rate. In
principle, there could be some discontinuity which separates these two
regimes. However, we will demonstrate that the scaling behaviour changes smoothly. We will investigate this question in free bosonic and fermionic  field theories with time dependent masses, 
closely following our earlier work \cite{dgm1,dgm2,dgm3}. In particular,  we are able to exhibit Kibble-Zurek scaling in free scalar field analytically.  
Further we will also show, both analytically and numerically, that in the time interval $-\tKZ \lesssim t \lesssim \tKZ$ the time dependence of the one point function is indeed given by a scaling function, as described in eq.~(\ref{t_over_tkz}).

It is also interesting to analyze systems with a finite physical cutoff, like for instance, lattice models. In that case, we expect the universal fast scaling to be modified as the quench rate reaches the cutoff scale. In a future
communication we will investigate this effect in certain exactly solvable spin systems \cite{ddgms}.

The remainder of the paper is organized as follows: In Section \ref{past_res}, we briefly  review of some salient aspects of the scaling described above for fast smooth quenches and slow quenches. In Section \ref{exp_sols}, we derive exact expressions for the response to mass quench in free scalar and fermionic field theories with
protocols which are suitable to study at both slow and fast
quenches. In Sections \ref{results} and \ref{ECP}, we discuss the main results of this paper for a variety of different protocols. 
%regarding Trans and Cis-Critical Protocols and End-Critical Protocols, respectively.
We conclude with a brief discussion of our results in Section \ref{discuss}. Finally in Appendix \ref{subleading} we discuss subleading contributions to KZ scaling.

\section{Review of past results}
\label{past_res}
In this section, we review salient features of both fast and slow
quenches.  

\subsection{Fast but smooth quenches} \label{faster}
In \cite{dgm1,dgm2,dgm3}, we described the evolution of the expectation
value of various operators under a fast but smooth quench, as summarized
above --- see also the discussion in \cite{david}. Note that for $\Delta >d/2$, the expectation
value $\vev{\cO_\Delta }_{ren}$ in eq. (\ref{0-4}) diverges as $\dt \to 0$. This result appears
paradoxical at first sight, since at least in low dimensions, there
are perfectly reasonable results for truly instantaneous quenches \cite{cc2,cc3,cc4}. This issue was examined in detail in \cite{dgm3} by considering UV finite quantities, such as correlation functions at finite spatial
separations and the excess energy produced. Again, the key difference between these instantaneous quenches and the present quenches is the relation between the quench rate $1/\dt$ and the UV cutoff $\Lambda_\mt{UV}$. In particular, for the fast but smooth quenches, we maintain $\Lambda_\mt{UV}\gg1/\dt$, as indicated in eq.~\reef{0-5}. We refer the interested reader to \cite{dgm3} for a thorough discussion.  Below, we review the most important results for the fast but smooth quenches that will be used in subsequent sections.

\begin{table}[]
\centering
\label{my-label}
\begin{tabular}{|c|c|c|c|}
\hline
Quench Type & Coupling & Operator         & Dimension    \\ \hline \hline
Fermions    & $m(t)$   & $\bar{\psi}\psi$ & $\Delta=d-1$ \\ 
Scalars     & $m^2(t)$ & $\phi^2$         & $\Delta=d-2$ \\ \hline
\end{tabular}
\caption{Description of free field theory quenches.}
\end{table}
As described above, the scaling in eq. (\ref{0-4}) is quite general but here we will focus on quenches for free scalars and free Dirac fermion fields with time dependent mass terms. The parameters characterizing these free field quenches are given  in Table \ref{my-label}. Hence in the limit of fast but smooth quenches, eq.~\reef{0-4} becomes
\begin{eqnarray}
\vev{\bar{\psi}\psi}_{ren} & \sim & {m}/{\dt^{d-2}} \, ,\label{drool1}\\
\vev{\phi^2}_{ren} & \sim & m^2 / \dt^{d-4} \,.\labell{drool2}
\end{eqnarray}
In \cite{dgm1,dgm2,dgm3}, we demonstrated that the above scaling holds both numerically and analytically. 

In particular, for an odd number of spacetime dimensions, we found that the leading order response  was given by
\begin{eqnarray}
\langle \bar{\psi}\psi \rangle_{ren} & = &  (-1)^{\frac{d-1}{2}}\, \frac{\pi}{2^{d-1}\,\sigma_f}\, \partial_t^{d-2} m(t/\dt) + O(\dt^{1-d})\qquad {\rm for}\ d\ge3\,,
\label{boat} \\
\langle\phi^2\rangle_{ren} & = & (-1)^{\frac{d-1}{2}}\, \frac{\pi}{2^{d-2}\,\sigma_s}\, \partial^{d-4}_t m^2(t/\dt) + O(\dt^{3-d})\qquad {\rm for}\ d\ge5\,,
\label{phi_odd}
\end{eqnarray}
where $\sigma_f$ and $\sigma_s$ are constants that only depend in
the spacetime dimension $d$ --- see eqs. (\ref{sigmaff}) and (\ref{sigs}). 
In an even number of spacetime dimensions, there is an enhancement of the scaling due to an extra logarithmic divergence in the counterterms. The leading response in the free field quenches then becomes \citep{dgm2}
\begin{eqnarray}
\langle \bar{\psi}\psi \rangle_{ren} & = &  (-1)^{\frac{d}{2}-1}\, \frac{\log(\mu\dt)}{2^{d-2}\,\sigma_f}\, \partial_t^{d-2} m(t/\dt) + O(\dt^{2-d})\qquad {\rm for}\ d\ge4\,,
\label{boat2} \\
\langle\phi^2\rangle_{ren} & = & (-1)^{\frac{d}{2}}\, \frac{\log(\mu\dt)}{2^{d-3}\,\sigma_s}\, \partial^{d-4}_t m^2(t/\dt) + O(\dt^{4-d})\qquad {\rm for}\ d\ge6\,,
\label{phi_odd2}
\end{eqnarray}
where $\mu$ is an additional renormalization scale. As a practical matter, this logarithmic enhancement was not very strong in our numerical calculations and it was important to account for the nonuniversal contribution appearing at the next order \citep{dgm2}. 

\subsection{Kibble-Zurek scaling}
\label{kzphysics}

 We now turn to a quench which is slow compared to physical scales in the problem. 
This is the regime where one expects Kibble-Zurek scaling \cite{kibble,zurek,more,gritsev,scfn,qcritkz}.
Generally,  this would mean that we will be in the adiabatic regime and so intuitively, the expectation value is just that corresponding to a fixed-mass expectation value with the mass at that particular instant of time.\footnote{This intuition is inaccurate when the spacetime dimension is sufficiently large because counterterms include contributions involving time derivatives of the mass profile \cite{dgm1,dgm2,dgm3} --- see eqs.~\reef{ct_f} and \reef{ict}.}
However, as noted in the introduction, that is no longer true if the quench
involves a critical point. In this case, it is impossible to maintain the adiabatic evolution in the vicinity of the critical point.
In particular, adiabatic perturbation theory will break
down when the  change in the instantaneous gap $E_{gap}(t)$ becomes of the same order
as the gap itself, \ie
\ben
\frac{1}{E_{gap} (t)^2}\, \frac{d E_{gap}(t)}{d\,t}  \simeq 1 \,.
\label{1-1}
\een

Now consider a quench where the time dependence of the coupling is described by  eq.~(\ref{0-1}) near the critical point. The instantaneous gap is given by \footnote{We are dealing with relativistic theories for which the dynamical critical exponent $z=1$.}
\ben
E_{gap} (t) \simeq  | g(t) - g_c |^{\nu} \,,
\label{1-2}
\een
and it then follows from eq. (\ref{1-1}) adiabaticity breaks down at the Kibble-Zurek time, $\tKZ$:
\ben
\tKZ \simeq \left( \frac{g_0}{\dt^{\,r}} \right)^{-\frac{\nu}{r \nu +1}} \,.
\label{1-3}
\een
Kibble and Zurek assumed that the system switches to a diabatic evolution between $t = -\tKZ$ and $t= \tKZ$, for symmetric protocols that cross the critical point at $t=0$.
If one further assumes that $\tKZ$ defines the only relevant physical scale during this period, the scaling for the expectation values can be determined by dimensional analysis, \ie
\ben
\vev{\calo_\Delta}_{ren} \simeq \frac{1}{\tKZ^{\,\Delta}} \,.
\label{1-4}
\een
Substituting eq.~(\ref{1-3}) in eq. (\ref{1-4}) then yields to eq.~(\ref{0-2}). Similar expressions have been predicted by Kibble and Zurek for the density of defects produced when sweeping across an order-disorder transition (see, for instance, \cite{qcritkz}).
Similar arguments extend the KZ scaling to a more precise description of the time period $|t|\lesssim\tkz$ with universal scaling functions \cite{scfn, qcritkz}. For example, the one-point function behaves as 
\ben
\vev{\cO_\Delta (t) }_{ren} \simeq \frac{1}{\tKZ^{\,\Delta}} \,F(t/\tkz) \,.
\label{t_over_tkz}
\een
Higher point correlation functions exhibit a similar scaling form  \cite{qcritkz}.

In the following, we will also consider protocols which approach the critical point with a exponential decay, \ie $|g(t)-g_c| \sim g_0 \exp (-t/\dt)$. In this case, while adiabaticity still breaks down at some finite time, the amount of time required to reach the critical point is always infinite. On the other hand, as measured by the energy scale of the gap, distance to the critical point remains finite. Hence it is more appropriate to define the Kibble-Zurek energy, $\EKZ$, as the value of the instantaneous gap when eq.~\reef{1-1} is satisfied. Following the analogous arguments as above, one then finds
\ben
\vev{\calo_\Delta}_{ren} \simeq \EKZ^{\,\Delta} \,.
\label{1-4X}
\een
Note that for a power-law profile as in eq.~\reef{0-1}, the KZ time and KZ energy are simply related by $\tKZ = 1/\EKZ$
and so eqs.~\reef{1-4} and \reef{1-4X} are identical in this case. However, in the exponential protocol, there is no such relation and we will show that eq.~\reef{1-4X} gives the correct scaling in this case in section \ref{ECP}.

In sections \ref{exp_sols} and \ref{results}, we will consider quenches in free field theories where the gap is linear in time 
near the critical point, \ie $E_{gap}(t) \simeq\frac{m}{\dt}\,t$.  This then defines the Kibble-Zurek time as
\beq
\tKZ = \sqrt{\dt/m} \,. \labell{ttkz}
\eeq
%Often it will be convenient to work with dimensionless time and so we define
%\ben
%\tau_\mt{KZ} \equiv \frac{\tKZ}{\dt} = \frac1{\sqrt{m \dt}}\,.
%\labell{dimen}
%\een
For the free field quenches, eq.~\reef{1-4} then yields
\begin{eqnarray}
\langle \bar{\psi}\psi \rangle_{ren} & \simeq &\frac1{\tkz^{d-1}} = \left(\frac{m}{\dt} \right)^{\frac{d-1}{2}} \,,
\labell{KZscalingTCP}\\
\vev{\phi^2}_{ren} & \simeq &\frac1{\tkz^{d-2}}= \left(\frac{m}{\dt} \right)^{\frac{d-2}{2}}  \,. \labell{KZscalingCCP}
\end{eqnarray}

In section \ref{ECP}, we will consider an exponential approach to the critical point in scalar field quenches, \ie $E_{gap}(t) \simeq m\,\exp(-t/\dt)$. In this case, $\EKZ = 1/\dt$ and hence eq.~\reef{1-4X} yields
\ben
\vev{\phi^2}_{ren} \simeq \frac{1}{\dt^{d-2}}\,.
\label{KZscalingECP}
\een
Note that the scaling behaviour here does not depend on $m$, the initial amplitude of the gap.

\section{Explicit solutions}
\label{exp_sols}

\begin{figure}[H]
        \centering
        \subfigure[TCP for Fermionic Quench]{
                \includegraphics[scale=0.36]{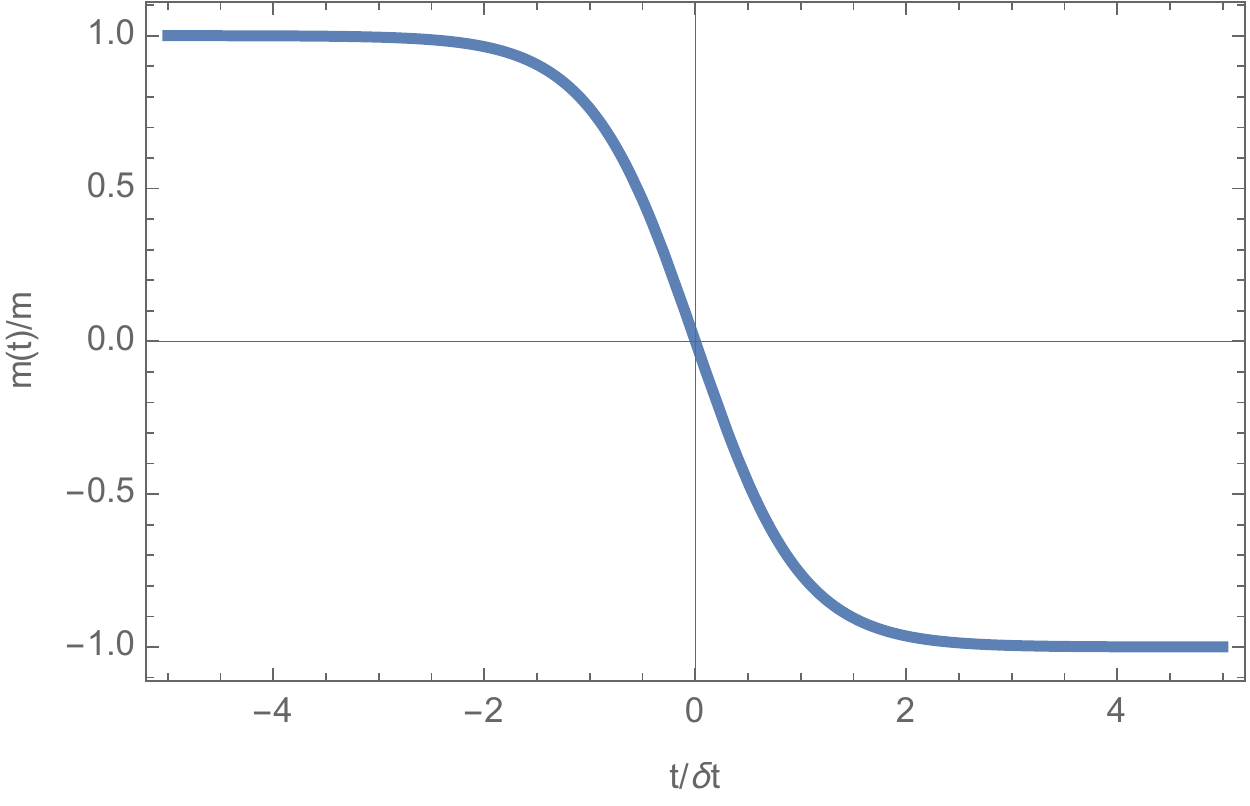} }
   		 \subfigure[CCP for Scalar Quench]{
                \includegraphics[scale=0.36]{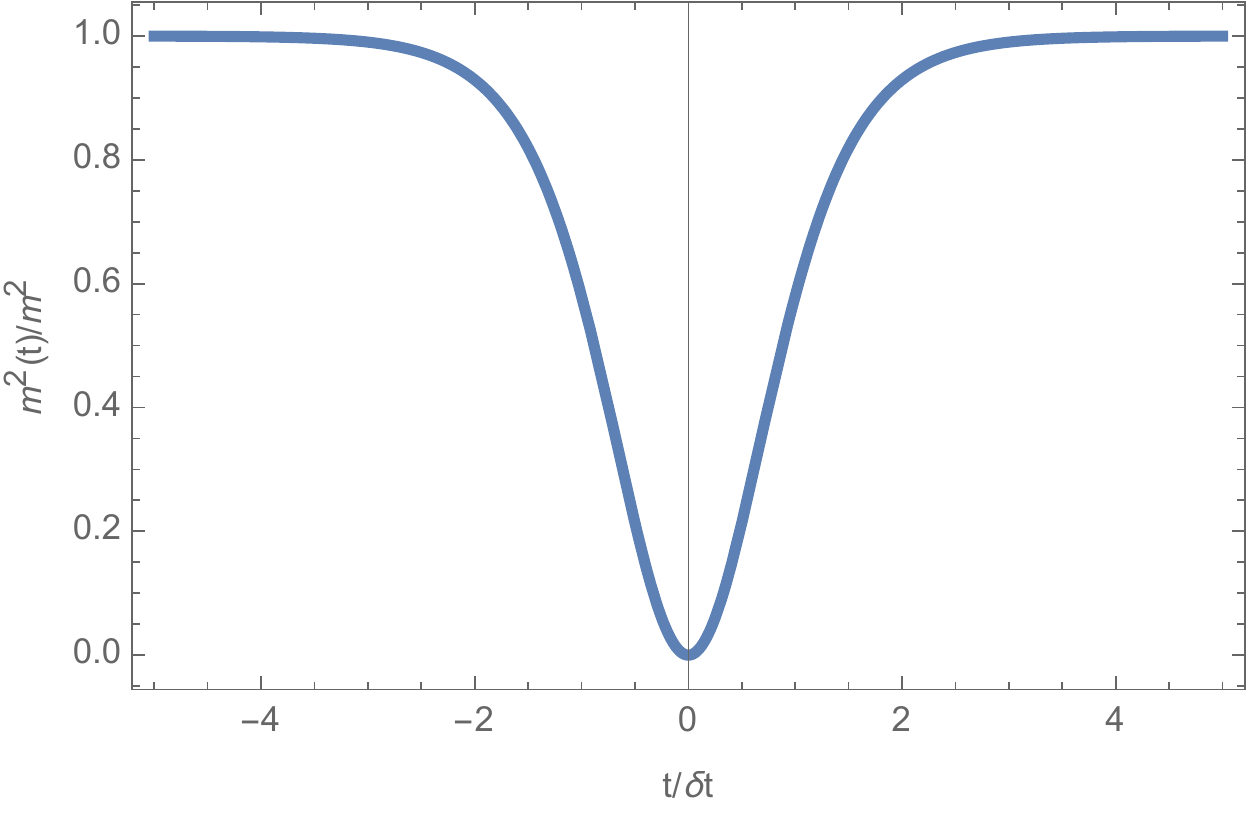} }
         \subfigure[ECP for Scalar Quench]{
                \includegraphics[scale=0.36]{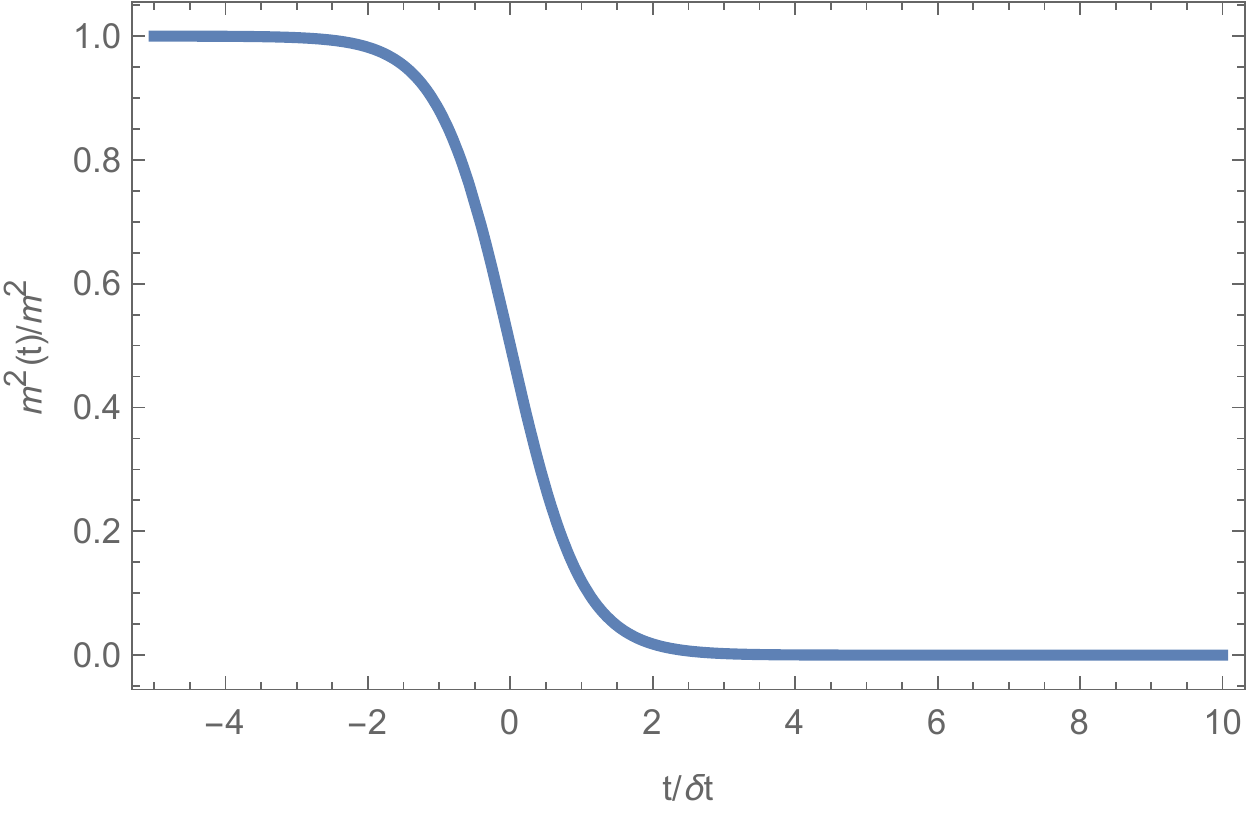} }
        \caption{Quench profiles to study KZ behaviour.} \label{fig_mass}
\end{figure}

Following \cite{qcritkz}, we will analyze three different protocols for which slow quenches should exhibit Kibble-Zurek scaling when approaching the critical point. As illustrated in figure \ref{fig_mass}, we will consider: Trans-Critical Protocols (TCPs), which cross through a critical point at $t=0$; Cis-Critical Protocols (CCPs), which only touch the critical point at $t=0$; and End-Critical Protocols (ECPs), which approach the critical point as $t\to\infty$.  In each case, we are considering a free scalar or fermion field and varying the mass. The key feature that makes our analysis manageable for free fields is that we are able to choose mass profiles for which we are able to solve the resulting field equation exactly and determine all of the modes analytically.

For the TCP, we consider a quench of the Dirac fermion where the mass varies with a profile proportional to $\tanh(t/\dt)$. In our previous work \cite{dgm1,dgm2,dgm3}, we were able to determine the modes for $\tanh$ profiles where the mass interpolates from $(A-B) m$ at $t\to-\infty$, to $(A+B) m$ at $t\to\infty$, with $A$ and $B$ being arbitrary constants. Our past investigations focused on quenches to the critical point with $A=-B$, and {\it reverse} quenches where $A=B$. In the present situation, we want the mass to go through zero at $t=0$, so we will simply choose $A=0$ and $B=-1$, \ie $m(t) = - m \tanh (t/\dt)$. Note that this means that the mass will be negative for $t>0$, however, this is not a problem in the fermionic field. 

For the scalar quenches, the coupling is the mass squared and  having a negative mass squared would result in unstable modes, \ie imaginary frequencies. 
To avoid this problem, for the scalar case, we will consider a CCP and will be analyzing pulsed quenches, where basically the mass squared will start at some positive value at $t=-\infty$, then go down to zero at $t=0$ and then return to that same positive value at $t=+\infty$. This analysis requires a slight extension of the study of pulsed mass profiles proportional to $1/\cosh^2(t/\dt)$ in \cite{dgm2,dgm3}. 

Finally, to consider the ECP, we will use the scalar field quench with a $\tanh$ profile which starts at some $m$ when $t=-\infty$ and approaches zero mass as $t\to\infty$, \ie $m^2(t)=\frac{m^2}2\left(1-\tanh (t/\dt) \right)$. Of course, this profile was already extensively studied in \cite{dgm1,dgm2,dgm3}, however, there we focused on the scaling behaviour near $t=0$ in the regime $m\dt\ll 1$. In the present case, we will consider the late time behaviour where $m^2(t)\simeq m^2 \exp\left[- 2t/\dt\right]$. As emphasized above, with this exponential approach to the critical point, our analysis of the ECP differs from that discussed in \cite{qcritkz} which considered ECP protocols with a power-law decay.

Again, sketches of the three different profiles are shown in figure \ref{fig_mass}. Note that the first two protocols need some changes from the original solutions given in \cite{dgm1,dgm2,dgm3}, and hence in the next subsections we will describe the required analysis of both quenches to critical points in some detail. For completeness, we will also show the mode solutions and calculation of the corresponding expectation value for the ECP, however, these can already be found in \cite{dgm1,dgm2,dgm3}.

\subsection{Trans-Critical Protocol for fermionic quenches}
We start by describing the exact mode solutions for a free fermionic quench with a TCP protocol. More general solutions for a mass profile of the form $m(t)=A+B \tanh(t/\dt)$, with arbitrary $A$ and $B$ was already discussed in \cite{dgm2}. So here, we just need to specify convenient values for these parameters to produce the protocol of interest. In particular, we choose $A=0$ and $B=-1$ to produce the mass profile
\beq
m(t) = - m\, \tanh (t/\dt) \,,
\label{fermionquench}
\eeq
which crosses the critical point at $t=0$ with $m(t)\simeq - m\,t/\dt$. Hence following the discussion in section \ref{kzphysics}, we will expect KZ scaling behaviour to appear in the region $|t|\lesssim \tKZ = \sqrt{\dt/m}$ when $m\dt\gg1$.

Again, the evaluation of the mode solutions and the expectation value $\vev{\bar{\psi} \psi}$ can be found in \cite{dgm2}.  
%\blu{The action is given by
%\beq
%S = \int dt\int d^{d-1}\vx~\bar{\psi} \left[ \gamma^\mu\partial_\mu - m(t) \right] \psi
%\eeq}
For the specified values of $A$ and $B$, the expectation value is given by
\begin{eqnarray}
\langle \bar{\psi}\psi \rangle & = & \sigma_f^{-1} \int \psi_{div}(k)\, dk  = \sigma_f^{-1} \int k^{d-4} dk \left( \frac{\omega+m}{2 \omega}\right) \nonumber \\
& &\quad \times \Big[ \left( k^2 - m^2(t) \right) |\phi_\vk|^2  - |\partial_t \phi_\vk|^2 + 2 m(t)\, \text{Im} \left( \phi_\vk\, \partial_t \phi_\vk^*\right) \Big]\,,
\label{barpsipsi}
\end{eqnarray}
where $\omega = \sqrt{k^2+m^2}$ and the functions $\phi_{\vk}$ are given by
\begin{eqnarray}
\phi_{\vk} (t) & = &  \exp \left( -i \, \omega \, t \right) \, _2F_1 \left(1 - i \, m \, \dt, i \, m \, \dt; 1- i \, \omega\, \dt; \frac{1+\tanh t/\dt}{2} \right) \, ,
\label{phi_fermions}
\end{eqnarray}
where $_2F_1$ is the usual hypergeometric function. Further, $\sigma_f$ is a numerical coefficient that depends on the spacetime dimension $d$,
\beq
\sigma_f = \left \{
\begin{array}{ll}
2^{1-d/2}\,(2\pi)^{\frac{d-1}{2}} / \Omega_{d-2} & \ \ {\rm for\ even}\ d\,, \\
2^{(3-d)/2}\, (2\pi)^{\frac{d-1}{2}} / \Omega_{d-2}& \ \  {\rm for\ odd}\ d\,,\ 
\end{array} \right.
\label{sigmaff}
\eeq
where $\Omega_{d-2}\equiv 2 (2\pi)^{(d-1)/2}/\Gamma((d-1)/2)$ is the solid angle in $d-1$ spatial dimensions.

As discussed in \cite{dgm1,dgm2,dgm3}, eq.~\reef{barpsipsi} is the bare expectation value, which is UV-divergent. The renormalized expectation value requires that we take into account the contributions for various counterterms and the final result takes the form
\beq
\langle \bar{\psi}\psi \rangle_{ren} \equiv \sigma_f^{-1} \int dk \Big[\psi_{div}(k) - f_{ct} (m(t),k) \Big]\,,\label{pppp}
\eeq
where $f_{ct}$ subtracts all the divergent terms in $\psi_{div}$ as $k\to \infty$. The counterterm contributions are given by \cite{dgm1,dgm2,dgm3},
\begin{eqnarray}
f_{ct}(m(t),k) & = & - m(t)\, k^{d-3} + \frac{m(t)^3}{2}\, k^{d-5} -  \frac{3 m(t)^5}{8}\, k^{d-7}+  \frac{1}{4} \partial^2_t m(t)\, k^{d-5} \labell{ct_f} \\
& &\qquad  - \left( \frac{1}{16}\, \partial^4_t m(t)  + \frac{5 m(t)}{8} \Big( \partial_t m(t) \partial_t m(t)  + m(t) \partial^2_t m(t)\Big) \right) k^{d-7}+\cdots , \nonumber
\end{eqnarray} 
where for a given value of $d$, one includes the terms where the power of $k$ is greater than or equal to $-1$.
Hence eq.~\reef{ct_f} includes all the necessary terms needed to regulate the fermionic expectation value \reef{pppp} up to $d=7$.  As we discussed in \cite{dgm1,dgm2,dgm3}, the first three terms are those needed to regulate the expectation value for a constant mass, while the remainder are novel contributions that involving time derivatives of the mass profile. It is significant that these counterterm contributions are written in a universal form that can be applied for any profile $m(t)$ that is a smooth function of time. In fact, these terms are obtained by considering the adiabatic expansion and performing a large-$k$ expansion of the answer.

Finally if we consider the free fermion in an odd number of spacetime dimensions $d$ with a fixed mass $m$, the renormalized expectation value of the fermion bilinear  becomes
\beq
\vev{\bar{\psi} \psi}_{ren,fixed} = \sigma_f^{-1} \frac{\Gamma \left(1-\frac{d}{2}\right) \Gamma \left(\frac{d-1}{2}\right)}{2 \sqrt{\pi }}\, m^{d-1}\, \text{sgn}(m(t)) \,.
\label{ferm_adiab}
\eeq

\subsection{Cis-Critical Protocol for scalar quenches}

For a scalar field with a pulsed mass profile that just touches the critical point, we need to slightly extend the analysis for a CFT-to-CFT quench studied in \cite{dgm2}. In fact, we will be consider here a slightly more general profile
\beq
m^2(t) = m_0^2 - \frac{m^2}{\cosh^2 (t/\dt)}\,.
\label{scalarquench}
\eeq
For this profile the equation of motion reads,
\beq
\frac{d^2 u_\vk}{dt^2} + \left( k^2 + m_0^2 - \frac{m^2}{\cosh^2 t/\dt} \right) u_\vk =0\,.
\label{eom2}
\eeq
By making the necessary substitutions in the solutions of the CFT-to-CFT quench in \cite{dgm2}, we obtain the following ``in" solution to (\ref{eom2}),
\begin{eqnarray}
u_\vk & = & \frac{1}{\sqrt{4\pi}(k^2+m_0^2)^{1/4}} \frac{2^{i \sqrt{k^2+m_0^2}} y^\alpha}{E'_{1/2} E_{3/2} - E_{1/2} E'_{3/2}} \times \label{modes_scalar} \\ 
& & \times \left( E_{3/2} \ _2F_1 (a,b;\frac{1}{2};1-y) + E_{1/2} \sinh(t/\dt) _2F_1 (a+\frac{1}{2},b+\frac{1}{2};\frac{3}{2};1-y) \right), \nonumber 
\end{eqnarray}
where
\begin{eqnarray}
E_{c} = \frac{\Gamma(c) \Gamma(b-a)}{\Gamma(b) \Gamma(c - a)} \  & , &  \ E'_{c} = E_{c} (a \leftrightarrow b)\,,
\nonumber \\
a = \alpha + \frac{i \, \dt \, \sqrt{k^2+m_0^2}}{2} & , & \ \ \ b  =  \alpha - \frac{i\, \dt \, \sqrt{k^2+m_0^2}}{2}, \label{a and b} \\
\alpha  =   \frac{ 1 - \sqrt{1- 4 m^2 \dt^2} }{4} & , & \ \ \ y=\cosh^2(t/\dt) \,.\nonumber
\end{eqnarray}
To recover the mode solutions for the CFT-to-CFT quench studied in \cite{dgm2}, we need to replace $m^2\to-m^2$ and set $m_0^2=0$. Instead we are interested here in the CCP quench which is achieved by setting $m_0 = m$ in the above expressions. With this choice, the mass profile \reef{scalarquench} becomes simply
\beq
m^2(t) =m^2\,\tanh^2(t/\dt)\,.
\label{scalarquenchX}
\eeq
In the vicinity of the critical point, \ie near $t=0$, we have $m^2(t) =m^2\,(t/\dt)^2$ and hence (away from $t=0$) the time dependence of the mass is again linear. Hence we again expect KZ scaling  to appear in the region $|t|\lesssim\tKZ = \sqrt{\dt/m}$ when $m\dt\gg1$.

In this case, we are interested in the expectation value of the mass operator $\phi^2$. Again, the bare expectation value is UV divergent but after introducing the necessary counterterms, a renormalized quantity becomes \cite{dgm1,dgm2,dgm3}:
\beq
\langle \phi^2 \rangle_{ren} = \sigma_s^{-1} \int dk \left( k^{d-2} |u_\vk|^2 - f_{ct} (k,m(t)) \right)\,,
\labell{reno}
\eeq
where
\beq
\sigma_s\equiv{2\,(2\pi)^{d-1}}/{\Omega_{d-2}}\,.
\label{sigs}
\eeq
The counterterm contributions can be obtained using an adiabatic expansion, as in \cite{dgm1,dgm2}. For completeness, we write the results needed to regulate the theory up to $d=9$,
\begin{eqnarray}
f_{ct} (k,m(t)) & = &  k^{d-3} - \frac{k^{d-5}}{2}\, m^2(t) + 
\frac{k^{d-7}}{8} \left( 3m^4(t)+ \partial^2_t m^2(t)\right) \label{ict} \\
& & - \frac{k^{d-9}}{32} \left(10 m^6(t) + \partial^4_t m^2(t) + 10 m^2(t)\, \partial^2_t m^2(t)+5\partial_t m^2(t)\, \partial_t m^2(t)\right)  + \cdots \,. 
\nonumber
\end{eqnarray}

As above, we end this subsection by noting that if we consider the free scalar in an odd spacetime dimension $d$ and with a fixed mass $m$, the renormalized expectation value becomes
\beq
\vev{\phi^2}_{ren,fixed} = \sigma_s^{-1} \frac{\Gamma \left(1-\frac{d}{2}\right) \Gamma \left(\frac{d-1}{2}\right)}{2 \sqrt{\pi }}\, m^{d-2} \,.
\label{sca_adiab}
\eeq

\subsection{End-Critical Protocol for scalar quenches}
\label{ECPsolution}

To study the ECP, we consider quenching the free scalar field quench with the following mass profile,
\beq
m^2(t)=\frac{m^2}2\left(1-\tanh (t/\dt) \right)\,.
\labell{ECPX}
\eeq
Such quenches were extensively studied in \cite{dgm1,dgm2,dgm3}, however, the focus was on the early-time scaling for fast quenches, \ie for $|t|\lesssim\dt$ with $m\dt\ll 1$. Here, we will examine these quenches for KZ scaling as we approach the critical theory, \ie for $t/\dt\gg1$ with $m\dt\gg1$. The late time behaviour of the above profile yields $m^2(t)\simeq m^2 \exp\left[- 2t/\dt\right]$. Hence following the discussion in section \ref{kzphysics}, KZ scaling behaviour should appear when $m(t)\lesssim \EKZ = 1/\dt$. Alternatively, we can phrase the latter as  $t\gtrsim\dt\log(m\dt)$.
 
For completeness, we exhibit here the exact mode solutions for these quenches, as shown in \cite{dgm1,dgm2,dgm3}.  Exact solutions to the Klein-Gordon equation with this mass profile is given by the following ``in" modes:
\begin{eqnarray}
u_\vk & = & \frac{1}{\sqrt{2 \omega_{in}}} \exp(i\vk\cdot\vec{x}-i\omega_+ t - i\omega_- \dt \log (2 \cosh t/\dt)) \times \label{inmodes} \\
& &\qquad_2F_1 \left( 1+ i \omega_- \dt, i \omega_- \dt; 1 - i \omega_{in} \dt; \frac{1+\tanh(t/\dt)}{2} \right)\,,
\end{eqnarray}
where 
$\omega_{in}  =  \sqrt{\vk^2+m^2}$, $\omega_{out}   = |\vk|$ and $\omega_{\pm}   =  (\omega_{out}  \pm \omega_{in})/2$.
As in the previous case, we compute the renormalized expectation value is given by eq.~\reef{reno}
with the counterterm contributions $f_{ct}$ appearing in eq.~(\ref{ict}).

\section{Results for TCPs and CCPs}
\label{results}
In the following, we will use the exact mode solutions described in the previous section to provide both analytic and numerical evidence that slow quenches going through a critical point exhibit KZ scaling near the critical point. In particular, we will study TCPs and CCPs in this section and leave the analysis of ECPs to the next section. In each case, we will be able to evaluate the corresponding expectation values at a fixed finite time, and by varying the quench rate, $1/\dt$, we will show a smooth transition from the fast quench regime to the the Kibble-Zurek regime and then to the adiabatic regime. The latter provides a complete display of the universal properties of quantum quenches at {\it any} rate within a single theory. 

\subsection{Breakdown of adiabaticity}
We start by showing that in fact there is some loss of the adiabatic behaviour that is manifest in our solutions near the critical point. To do this we fix a {\it large} value for $m \dt$ and we follow the evolution of the expectation values as a function of time. What we see is that, in general, the expectation value for the operator follows its adiabatic evolution, \ie eq.~(\ref{ferm_adiab}) for fermions and eq.~(\ref{sca_adiab}) for scalars with the mass given by the value of $m(t)$ at that particular time $t$. However, when the coupling approaches the critical point, we begin to see a deviation from the adiabatic evolution. In particular, in the interval set by the Kibble-Zurek time, \ie  $|t|\lesssim \tKZ$, the expectation value of the operator differs from the adiabatic result and it does not reach zero as the adiabatic answer would when $m=0$.

\begin{figure}[H]
        \centering
        \subfigure[Adiabatic behaviour]{
                \includegraphics[scale=0.35]{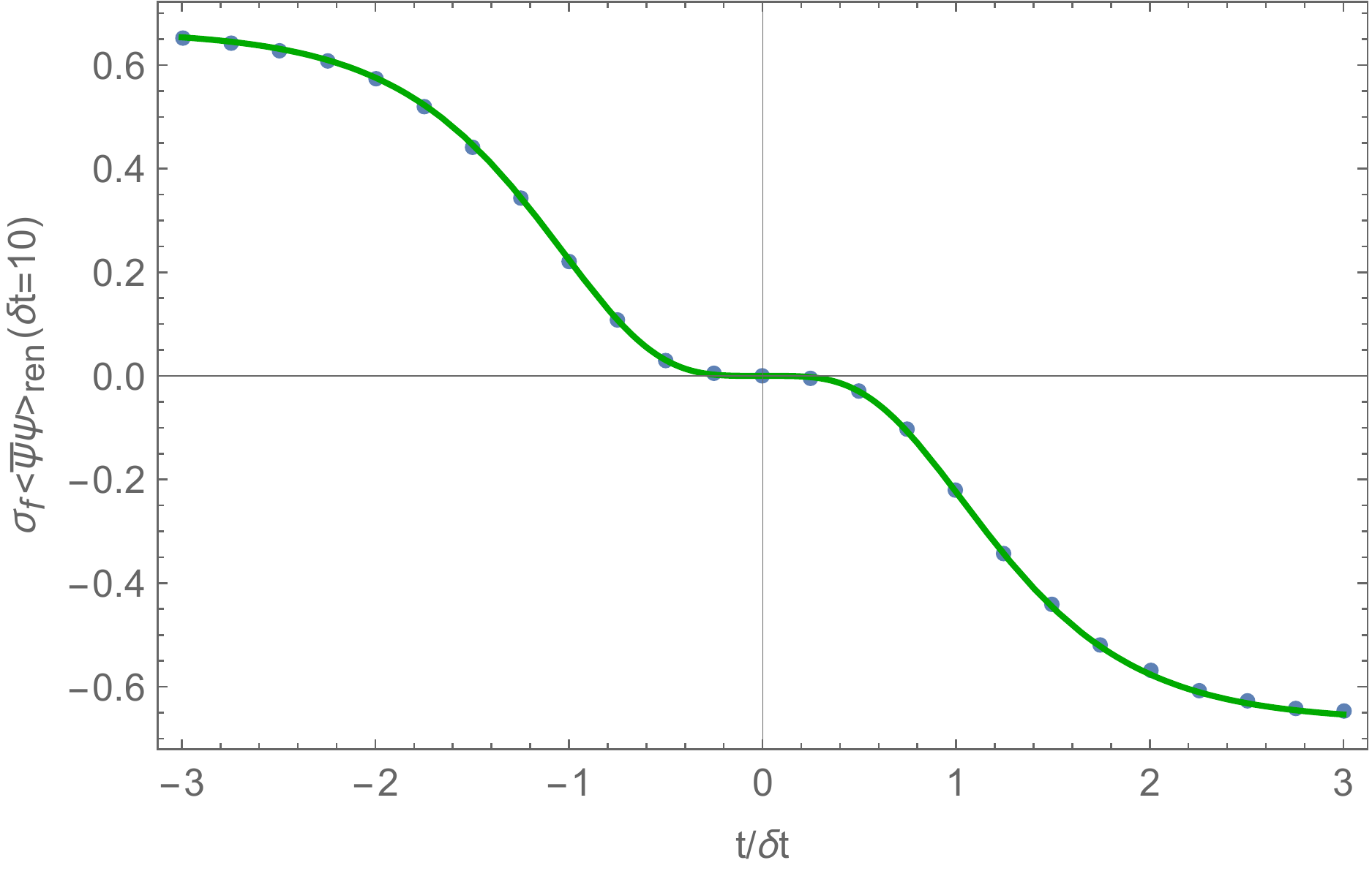} }
   		 \subfigure[Loss of adiabaticity near the critical point]{
                \includegraphics[scale=0.35]{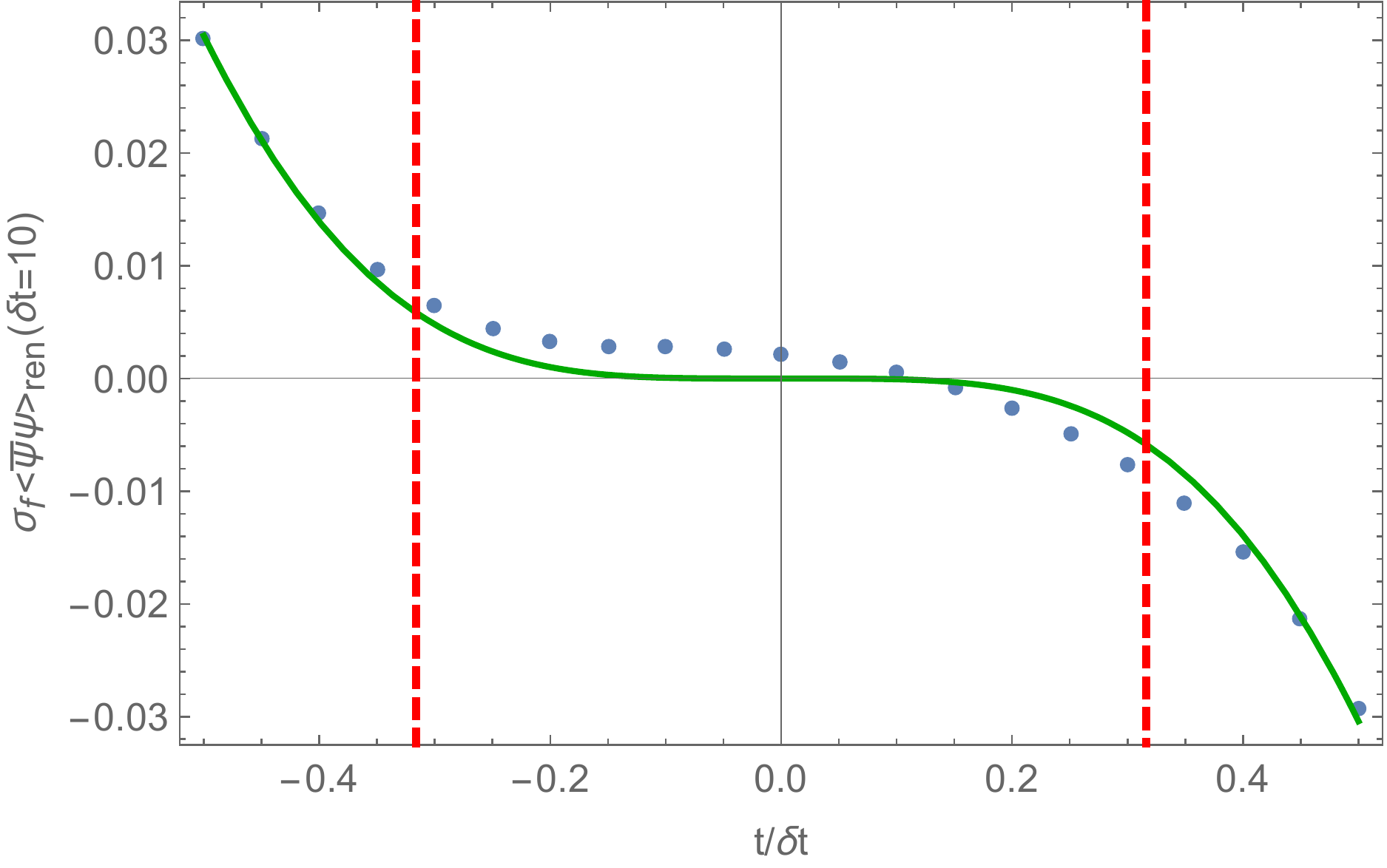} }
        \caption{Evidence for KZ physics near a critical point in TCP fermionic quenches. The green solid line represents the adiabatic solution  \reef{gulp}. The blue dots correspond to the exact expectation value \reef{pppp} of the mass operator for a slow quench with $m\dt =10$ and $d=5$. In panel (a), we see that for early and late times the expectation value follows the adiabatic expectation for slow quenches. In panel (b), we focus on the region near the critical point ($t=0$), and in fact, we see that the expectation value differs from the adiabatic one. As a guide we plotted in dashed red lines plus and minus the Kibble-Zurek time, $\pm \tkz/\dt= \pm 1/\sqrt{m \, \dt}$, where we should expect the two curves to start differing from each other, according to the original Kibble-Zurek argument. As we see in panel (b), this is in fact what is happening.
        } \label{fig_evidence_ferm}
\end{figure}

\begin{figure}[H]
        \centering
        \subfigure[Adiabatic behaviour]{
                \includegraphics[scale=0.35]{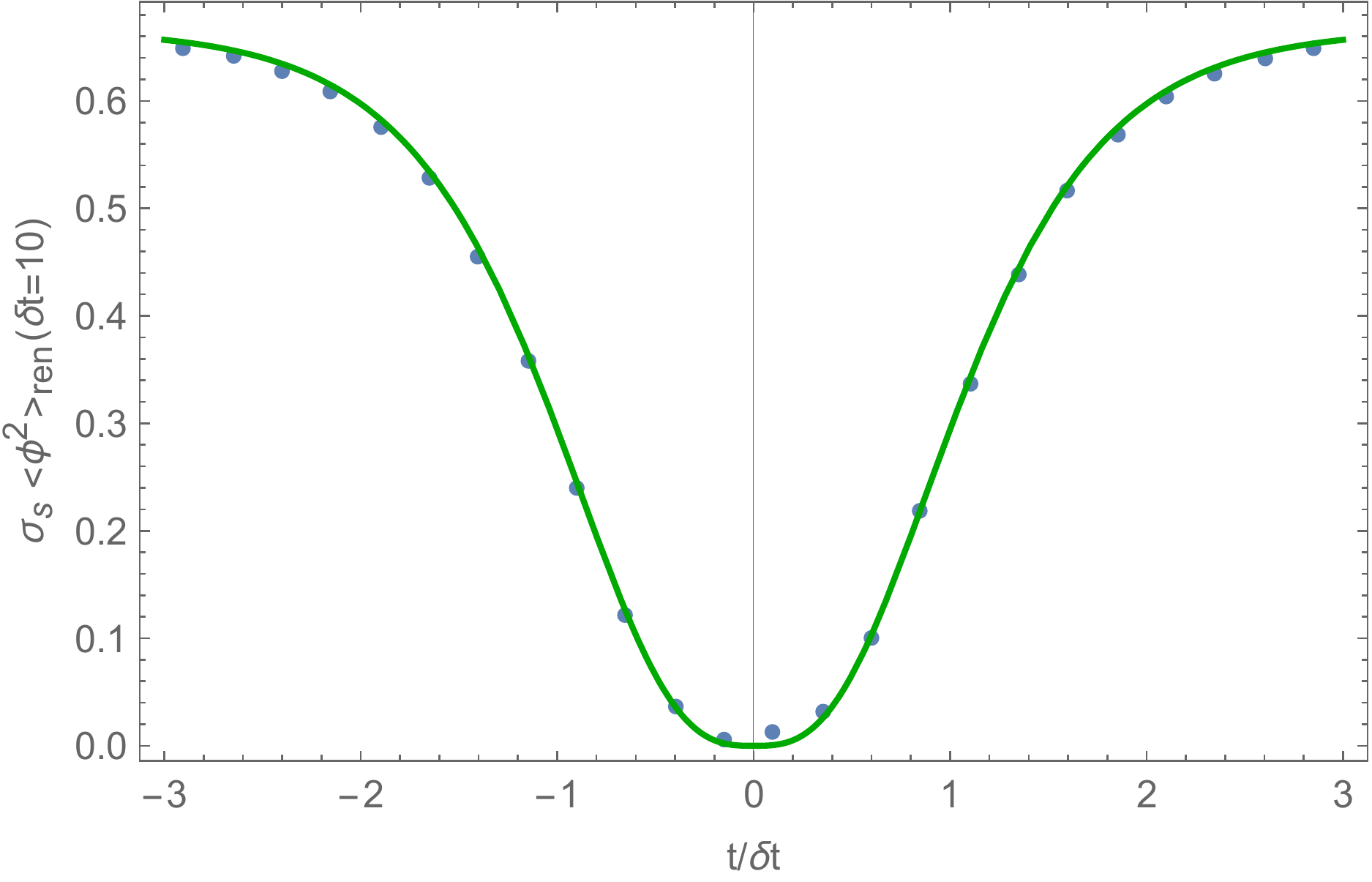} }
   		 \subfigure[Loss of adiabaticity near the critical point]{
                \includegraphics[scale=0.35]{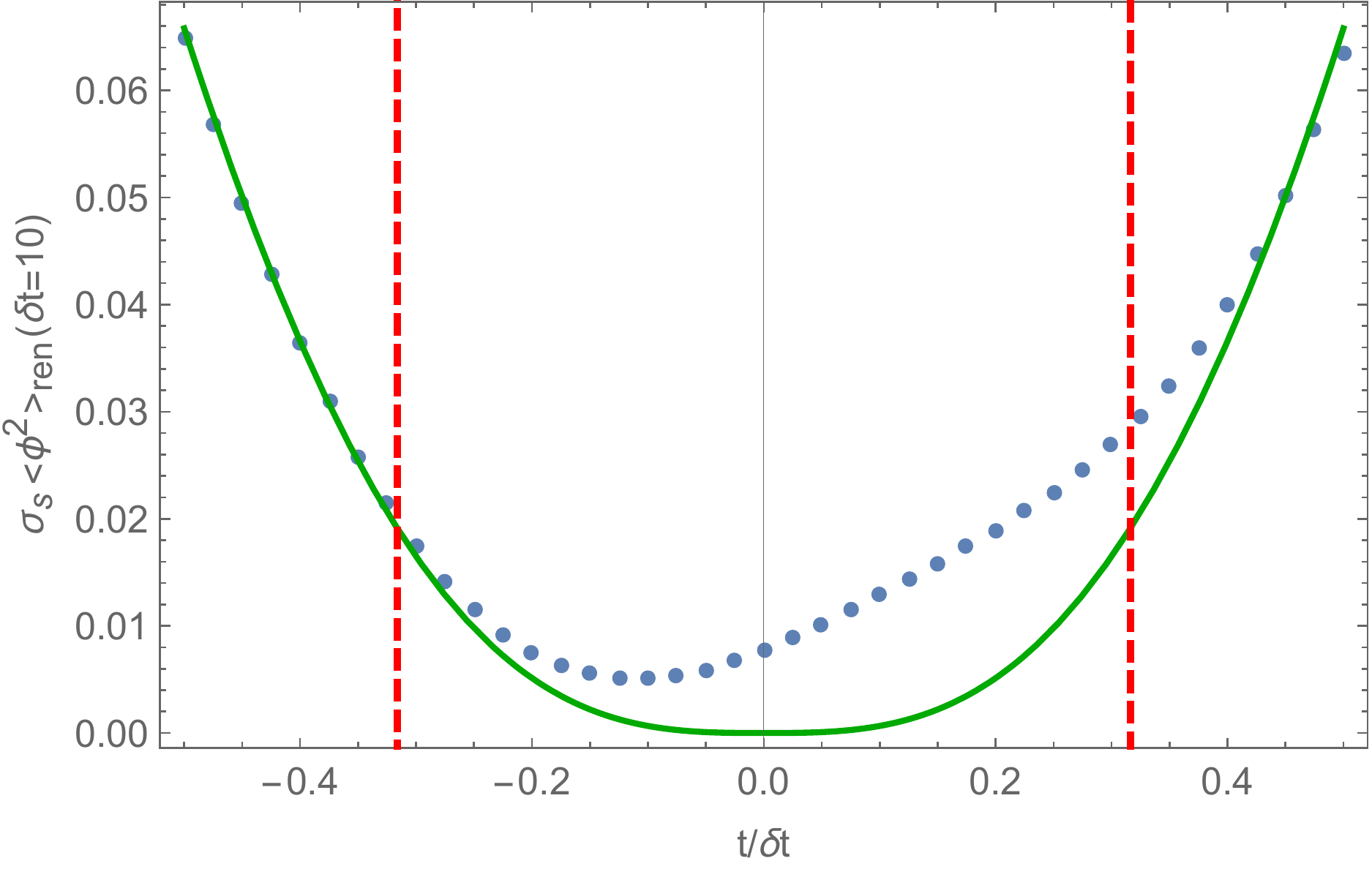} }
        \caption{Evidence for KZ physics near a critical point in CCP scalar quenches. The green solid line represents the adiabatic solution \reef{gulp2}. The blue dots correspond to the exact expectation value \reef{reno} of the mass operator for a slow quench with $m\dt =10$ and $d=5$.  In panel (a), we see that for early and late times the expectation value follows the adiabatic expectation for slow quenches. In panel (b), we focus on the region near the critical point ($t=0$), and in fact, we see that the expectation value differs from the adiabatic one. As a guide we plotted in dashed red lines plus and minus the Kibble-Zurek time, $\pm \tkz/\dt=  \pm 1/ \sqrt{m \, \dt}$, where we should expect the two curves to start differing from each other, according to the original Kibble-Zurek argument. As we see in panel (b), this is in fact what is happening.
        } \label{fig_evidence_sca}
\end{figure}

This general behaviour is illustrated for the TCP fermionic quenches in figure \ref{fig_evidence_ferm} and for the CCP scalar quenches in figure \ref{fig_evidence_sca}. In both cases, we show results for $\dt=10$, $m=1$ and $d=5$. For the fermionic case, eq.~\reef{ferm_adiab} yields the adiabatic solution as 
\ben
\sigma_f \vev{\bar{\psi}{\psi}}_{ren,adiabatic} = \frac{2}{3}\, m(t)^4\ {\text{sgn}} (m(t))\,,
\labell{gulp}
\een
and similarly, for the scalar quench, eq.~\reef{sca_adiab} produces 
\ben \sigma_s \vev{\phi^2}_{ren,adiabatic} = \frac{2}{3}\, m^3(t)\,.
\labell{gulp2}
\een

\begin{figure}[H]
        \centering
        \subfigure[Long times, $-4<t/\tkz<4$.]{
                \includegraphics[scale=0.39]{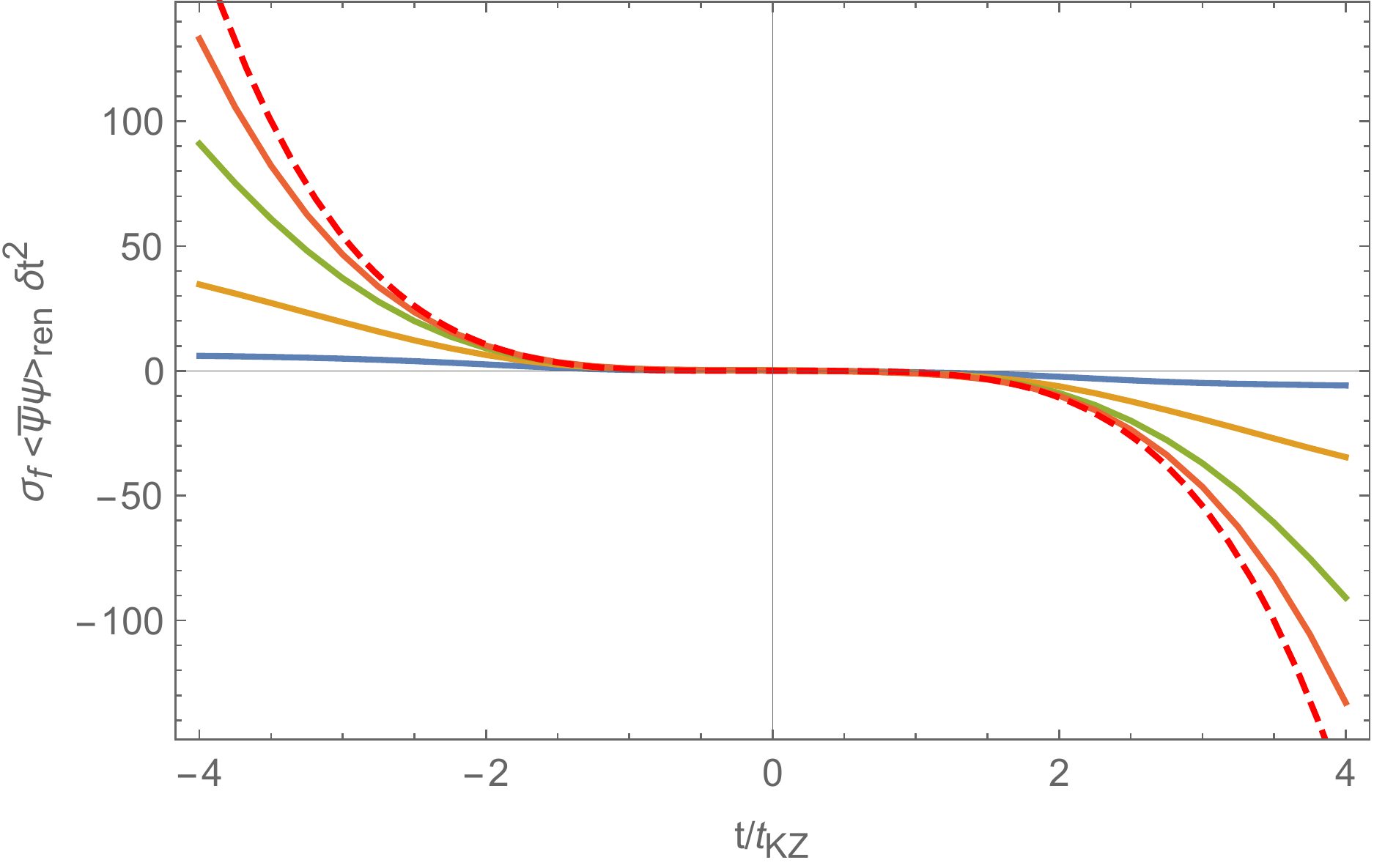} }
   		 \subfigure[Kibble-Zurek interval, $-1<t/\tkz<1$.]{
                \includegraphics[scale=0.39]{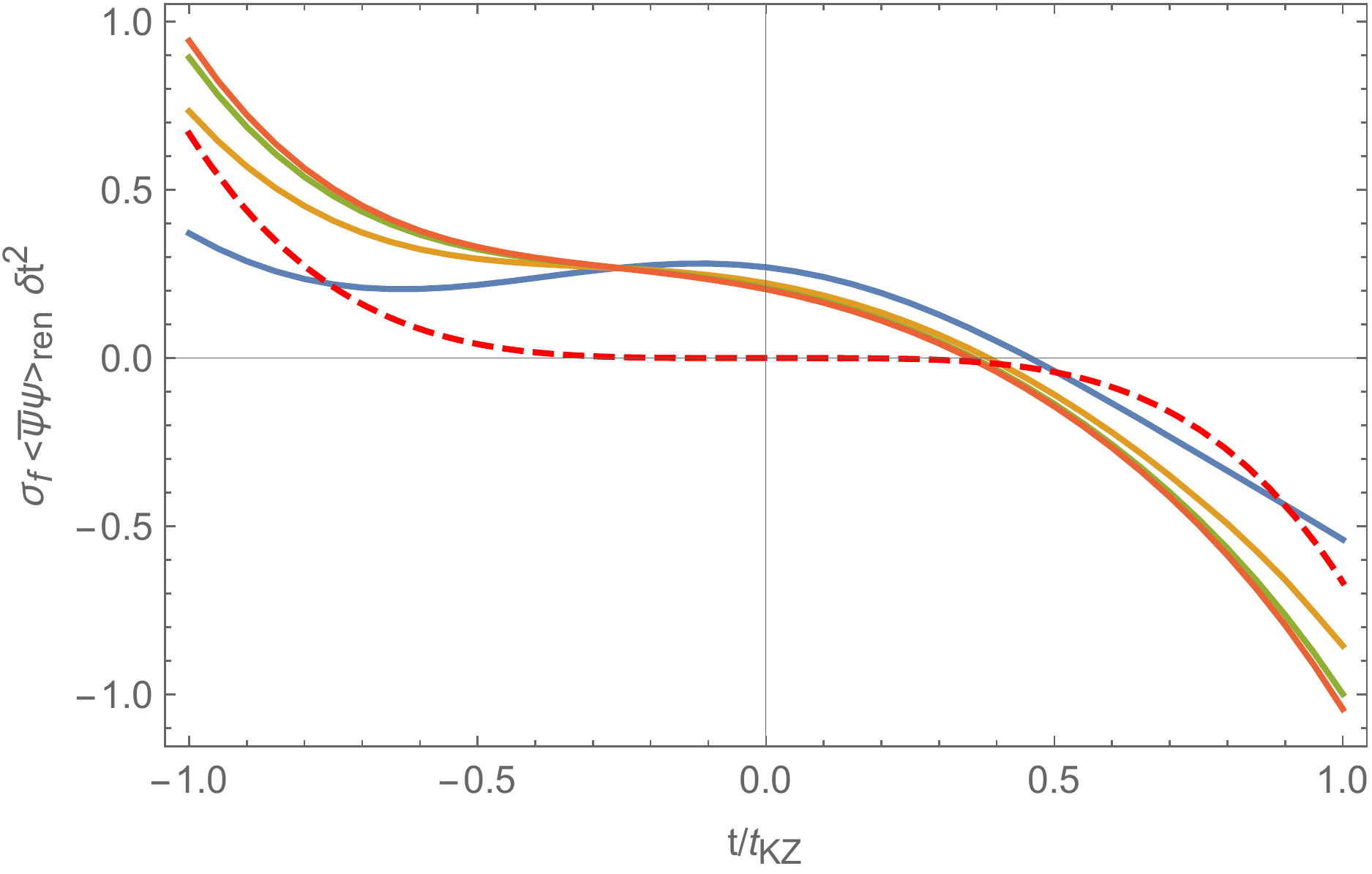} }
        \caption{Renormalized expectation value of fermionic mass operator as a function of $t/\tkz$ with $m=1$ and $d=5$. The different curves correspond to $\dt=10^{i}$, with $i=0.5 \text{(blue)},\ 1 \text{(yellow)},\ 1.5 \text{(green)},\ 2 \text{(orange)}$, in units of $m$. Note that we are multiplying the expectation value by $\dt^{2}$, that is the expected overall KZ scaling \reef{KZscalingTCP}. The dashed red curve plotted the adiabatic expectation value \reef{gulp}. In panel (a), we show the response for large time periods, while in panel (b), we zoomed in the interval where we expect KZ scaling to appear (\ie $-1<t/\tKZ<1$).
        } \label{fig_evidence_fer_2}
\end{figure}

As suggested by eq.~(\ref{t_over_tkz}), a useful way to observe the KZ scaling is by computing the renormalized expectation value of the quenched operator  as a function of $t/\tkz$. For symmetric protocols, then, we should expect the expectation value in the interval $|t|\lesssim \tKZ$  to be given by the overall KZ scaling factor times some function of $t/\tkz$. In figure \ref{fig_evidence_fer_2}, we plotted the expectation value for the mass operator in the TCP fermionic quenches for different quench rates $1/\dt$ as a function of $t/\tkz$ with $m=1$ and $d=5$. Note that we are extracting out the overall Kibble-Zurek scaling found in eq.~\reef{KZscalingTCP}, \ie we plot $\sigma_f \vev{\bar{\psi}{\psi}}_{ren}\, (\dt/m)^2$. As we increase $\dt$, we observe that (between $t/\tkz=\pm 1$) all the curves converge towards a single scaling function $F(t/\tkz)$, as in eq. (\ref{t_over_tkz}). Moreover, we plotted the adiabatic expectation value \reef{gulp} to show that outside of the KZ interval, the curves tend to approximate to the adiabatic one as $\dt \to \infty$. However, in the KZ region, the curves are clearly different from the adiabatic expectation. We computed analogous results for the CCP scalar quenches and in that case, we were also able to obtain an analytic result for $F(t/\tkz)$. So we reserve the discussion of this case for section \ref{anyrate}. However, the impatient reader can find the analogous plots in figure \ref{fig_evidence_sca_2}.

\subsection{KZ scaling of expectation values at $t=0$}
In order to characterize this special behaviour, we first concentrate on the expectation values at $t=0$. In this special case, the formulas are greatly simplified. For instance, all the counterterm contributions that are proportional to the mass vanish and so we do not need to consider them. As we will see below, at this particular time, we will be able to extract the KZ scaling analytically in the case of the CCP scalar quench. %This will also allow us to calculate the corrections to leading scaling.

\subsubsection{Numerical results}

We start by evaluating the expectation values numerically at $t=0$ for both the fermionic and the scalar quenches. This is, in principle, a challenging task because as we increase $\dt$, we expect the expectation value to approach zero. So in general we will be integrating numerically large quantities that will cancel to give a very small (and decreasing with larger $\dt$) number.

Another important aspect to note is that in fact the formulas presented in section \ref{exp_sols} are valid for any quench rate. In particular, for very small $\dt$, we should recover our past universal results for fast quenches (see section \ref{faster}), while for large $\dt$, we expect to find KZ behaviour. At this point, while focusing on $t=0$, the only (dimensionless) variable in the problem is $m \, \dt$ and so we expect the fast quench scaling to appear for $m \, \dt \ll 1$, and the KZ scaling, for $m \dt \gg 1$. Our exact expressions for the free field quenches also allow us to see the transition between these two regimes. 

Finally, note that at $t=0$, it is impossible to achieve adiabatic behaviour since the adiabaticity condition requires 
\ben
\frac{1}{m(t)^2}\frac{d m(t)}{dt} \ll 1 \,.
\een
This becomes, both for fermionic quenches with mass given by eq. (\ref{fermionquench}) and the pulsed scalar quench with mass given by eq. (\ref{scalarquenchX}),
\ben
m\dt \gg \frac{1}{\sinh^2(t/\dt)} \,,
\een
which can never be satisfied at $t=0$.

\begin{figure}[H]
        \centering
        \subfigure[$d=4$]{
                \includegraphics[scale=0.58]{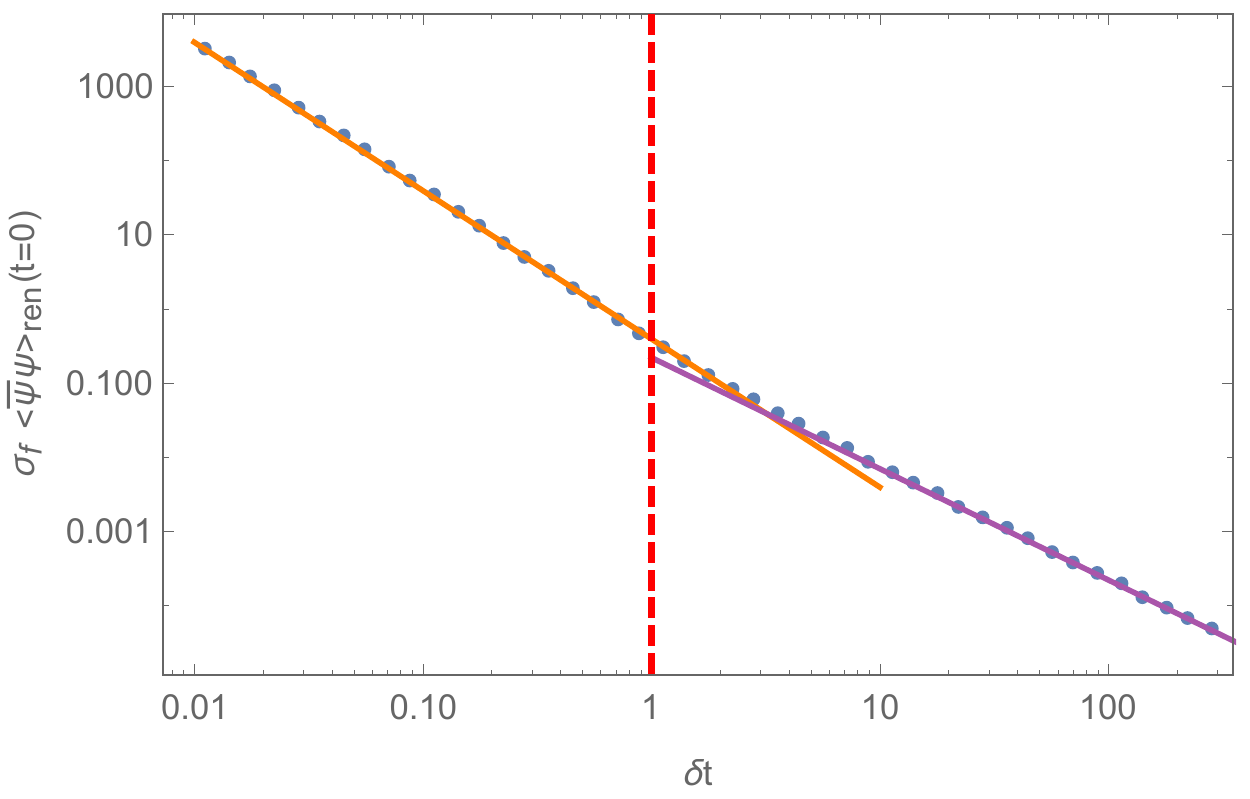} }
   		 \subfigure[$d=5$]{
                \includegraphics[scale=0.58]{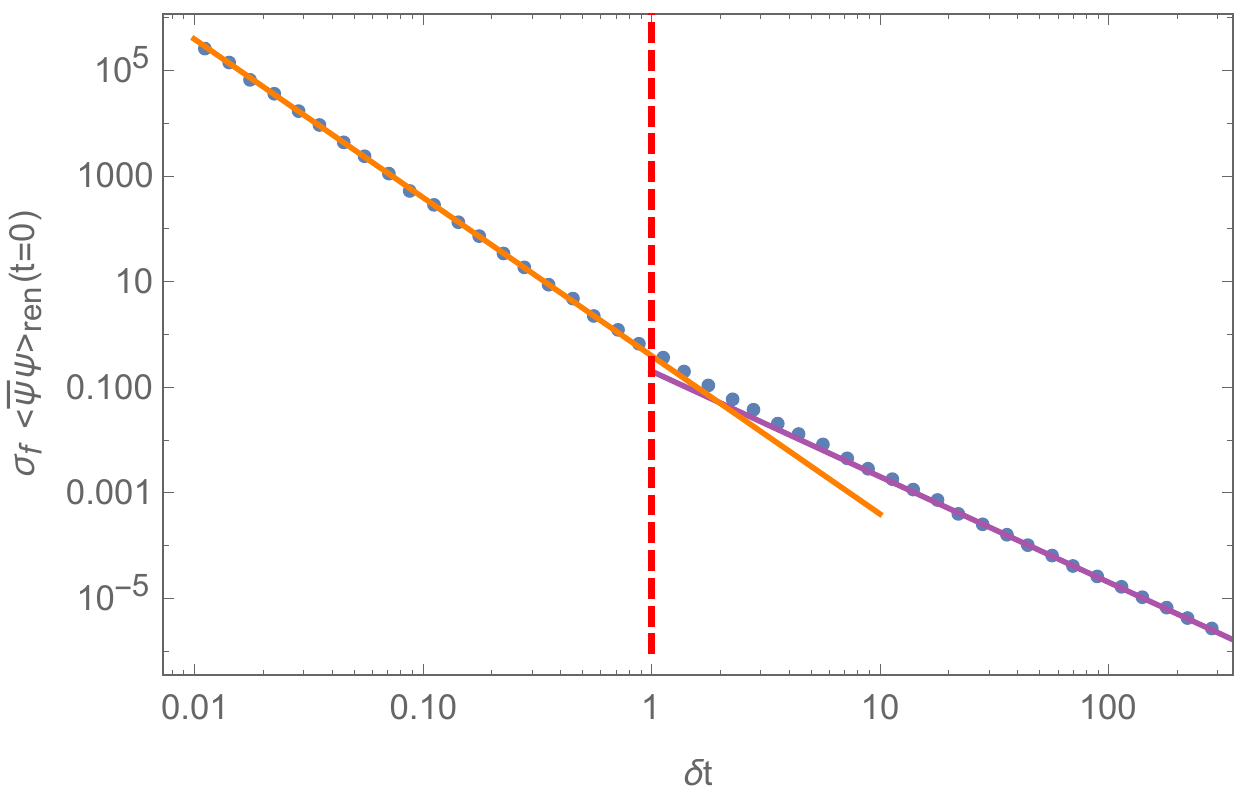} }
        \caption{The transition between the fast quench and slow quench regimes for expectation values at $t=0$ in the TCP fermionic quenches. The fast quench exhibits the usual fast quench universal scaling \reef{drool1}. The leading analytical contributions were found in \cite{dgm1,dgm2} and are plotted in solid orange. In solid purple, we have the best straight-line fits for the slow regime. The slope of these fits agrees with the expected KZ scaling, \ie $\vev{\bar{\psi}{\psi}}_{ren} \sim \dt^{-\frac{d-1}{2}}$. The two regimes are seperated by the scale $m \, \dt = 1$, which is plotted in dashed red as a guide to the eye only.} \label{fig_kz_zero_ferm}
\end{figure}

We start by analyzing the fermionic quench. We fixed the time to $t=0$ and then computed the expectation value for the mass operator for different values of $\dt$ for $d=4$ and $d=5$. The results are shown in figure (\ref{fig_kz_zero_ferm}). In the fast quench regime, we just reproduce the early-time scaling behaviour in eq.~\reef{drool1}, \ie $\vev{\bar{\psi}\psi}_{ren} \sim {m}/{\dt^{d-2}}$. In fact, our results give perfect agreement with the analytic expressions found in \cite{dgm1, dgm2}  (orange curves) --- see eqs.~\reef{boat} and \reef{boat2}. In the slow quench region, we find that the best fit curve reproduces the expected KZ scaling \reef{KZscalingTCP}, \ie $\vev{\bar{\psi}{\psi}}_{ren} \sim (m/\dt)^{\frac{d-1}{2}}$. In between these two regimes, \ie for $m\dt\sim 1$, we find a smooth transition between the two scaling behaviours. Also note that in the $d=4$ case, we do not find the logarithmic enhancement expected for even dimensions. This is because we are calculating the expectation value at $t=0$, where the logarithmic factor vanishes --- see equations (3.15) and (3.16) in \cite{dgm2}. In the next section, we will study expectation values at finite $t$ where we do expect to see this logarithmic enhancement in even dimensions.

\begin{figure}[H]
        \centering
        \subfigure[$d=4$]{
                \includegraphics[scale=0.38]{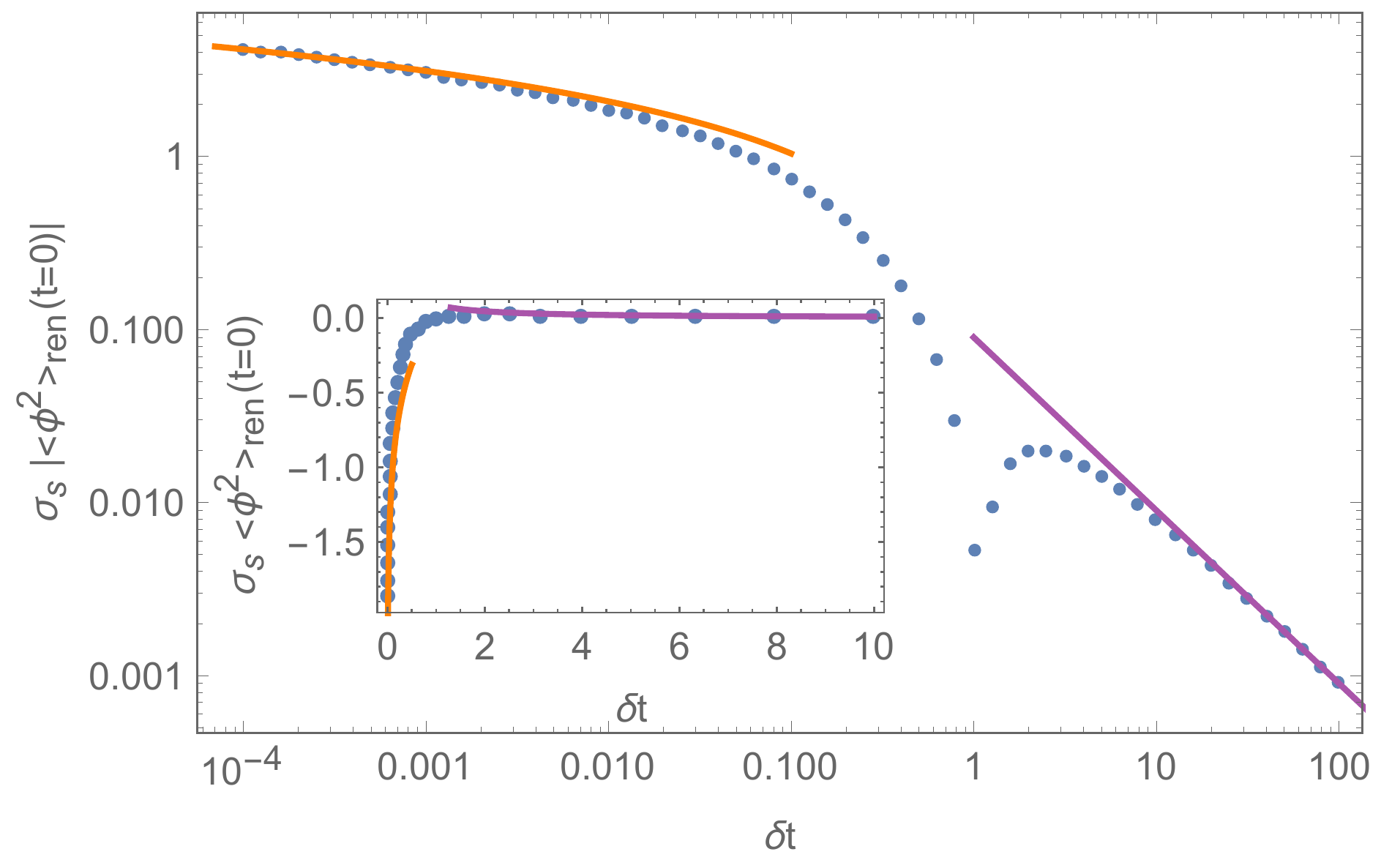} }
   		 \subfigure[$d=6$]{
                \includegraphics[scale=0.38]{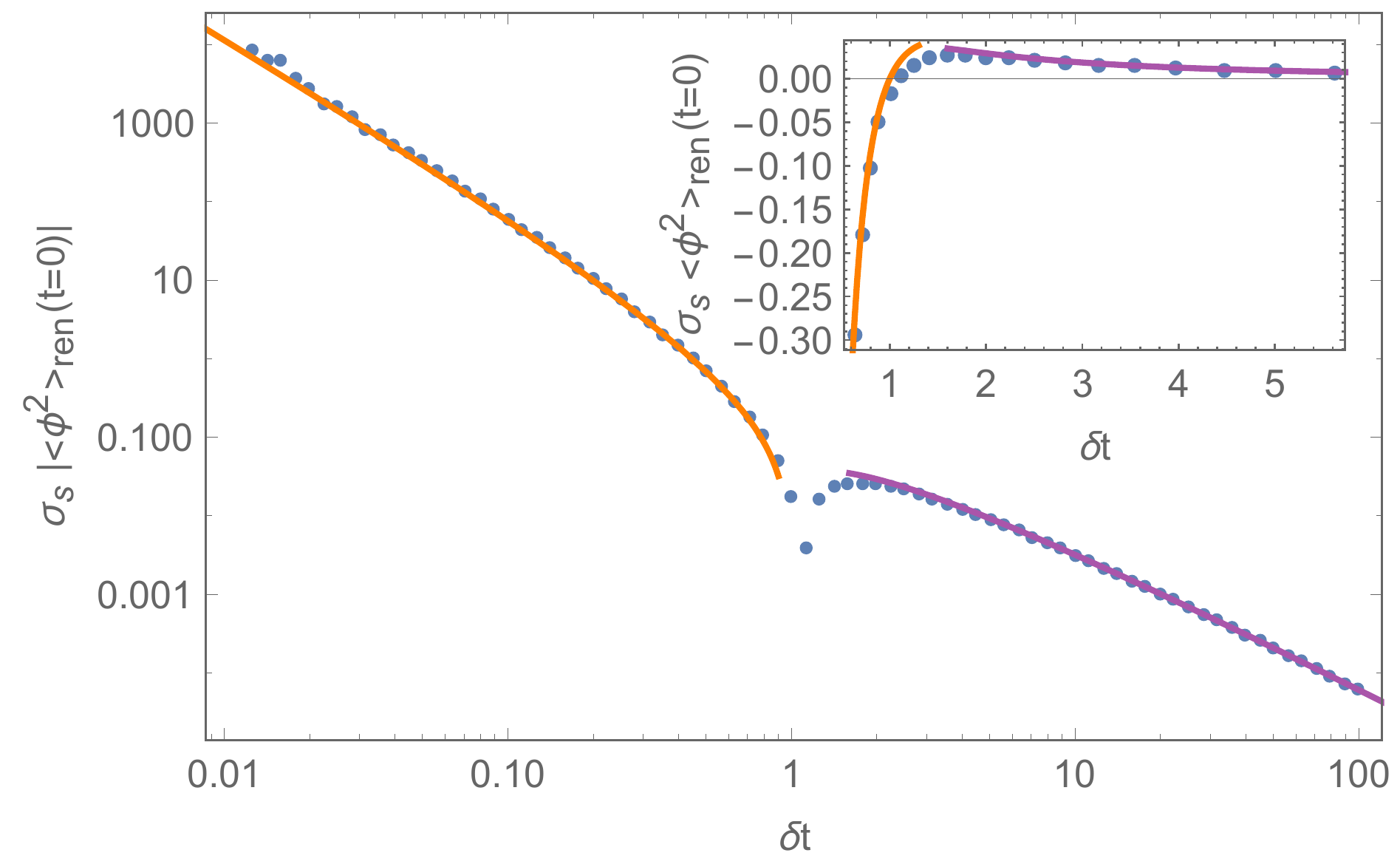} }
        \caption{The transition between the fast quench and slow quench regimes for expectation values at $t=0$ in the CCP  scalar quenches. The fast quench exhibits the usual fast quench universal scaling \reef{drool2}. The solid orange curve is the leading order contribution for fast quenches. This has an extra logarithmic factor as we are in even dimensions. The solid purple curves show the KZ scaling \reef{KZscalingCCP}, \ie $\vev{\phi^2}_{ren} \sim \dt^{-\frac{d-2}{2}}$, which is enhanced by a logarithmic factor in $d=6$ but not in $d=4$. In each figure, the inset shows the same expectation value but {\it{not}} on a logarithmic scale, near the region where the apparently singular behaviour appears. As shown there, the profiles are smooth and the apparent singular behaviour is just an artifact of the logarithmic scale when the expectation value changes sign.} \label{fig_kz_zero_sca}
\end{figure}

Now we turn to the case of free scalars with the mass profile given by eq.~(\ref{scalarquenchX}). In this case there is an extra feature: for fast quenches, the leading expectation value goes as the $(d-4)$'th derivative of the mass profile --- see eqs.~\reef{boat} and \reef{boat2}. This means that for odd dimensions, this leading contribution to $\vev{\phi^2}_{ren}$ vanishes at $t=0$. As a result, the fast scaling is very difficult to see, so we will focus our attention (for now) on even dimensions. The results can be seen in figure \ref{fig_kz_zero_sca}. For $d=4$, we see that for the fast quenches, there is a pure logarithmic scaling and when we transition to the slow quench, we find that it scales as $\dt^{-\frac{d-2}{2}}=\dt^{-1}$, that is the expected KZ scaling \reef{KZscalingCCP}. Note that in the intermediate region the expectation value changes sign and to continue plotting in the logarithmic scale we plot the absolute value of $\vev{\phi^2}_{ren}$. This generates apparently singular behaviour in the expectation value but that is just an artifact of the logarithmic scale, as can be seen from the insets in figure \ref{fig_kz_zero_sca}. There the profiles are plotted on a regular scale and we see the expectation value passes smoothly through zero.

Note that in $d=4$, the KZ scaling is not enhanced by a logarithmic factor, as it is in the fast quench regime. We will discuss this fact in the next section. In fact, this is special for $d=4$, because as can be appreciated in figure \ref{fig_kz_zero_sca}b. There we see a logarithmic enhancement in both the fast and the slow quench regimes for $d=6$, and the same holds for higher even dimensions. Apart from that difference, the behaviour and the two characteristic scalings are the same for $d=6$.

\subsubsection{Analytical results}
\label{ana_pulsed}
In order to get some analytical understanding of the quench process near the critical point, we study the pulsed scalar quench at $t=0$. A generalization of the ideas in this section is given in section \ref{anyrate} to evaluate the expectation value at finite times.

The expression for the quenched operator simplifies at $t=0$. Consider eq. (\ref{modes_scalar}) which we recall here,
\begin{eqnarray}
u_\vk & = & \frac{1}{\sqrt{4\pi}(k^2+m_0^2)^{1/4}} \frac{2^{i \sqrt{k^2+m_0^2}} y^\alpha}{E'_{1/2} E_{3/2} - E_{1/2} E'_{3/2}} \\ 
& & \times \left( E_{3/2} \ _2F_1 (a,b;\frac{1}{2};1-y) + E_{1/2} \sinh(t/\dt) _2F_1 (a+\frac{1}{2},b+\frac{1}{2};\frac{3}{2};1-y) \right). \nonumber
\end{eqnarray}
Now, at $t=0$, the second term of the second line vanishes because of the overall factor of $\sinh (t/\dt)$. Moreover, we remind the reader that $y=\cosh^2(t/\dt)$, so the last argument of both hypergeometric functions is $(1-y)|_{t=0} = 0$. This means that at $t=0$, the mode solutions are simplified  to
\begin{eqnarray}
u_\vk (t=0) = \frac{2^{i \sqrt{k^2+m_0^2}} }{\sqrt{4\pi}(k^2+m_0^2)^{1/4}} \frac{E_{3/2}}{E'_{1/2} E_{3/2} - E_{1/2} E'_{3/2}} \, .
\label{3-10}
\end{eqnarray}
Now we need to find the behaviour of this expression for large values of $m\dt$. Note that an adiabatic expansion would be a power series in $1/(m\dt)^{2}$ \cite{dgm2}, which is indeed a good expansion far from the critical point. However at $t=0$ adiabaticity has broken down, so this power series expansion is no longer valid.

The bare expectation value is given by
\ben
\vev{\phi^2}\big|_{t=0} = \sigma_s^{-1} \int dk~k^{d-2} |u_\vk (t=0)|^2 \,.
\een
An efficient way of extracting the large $m\dt$ behaviour is to make a change of variables in the above integral in a way which allows an expansion of the integrand for large $m\dt$. This is along the lines of the analysis of fast quench which was performed in \cite{dgm1,dgm2,dgm3} where we were looking for an expansion for small $m\dt$. In that case, it was useful to perform the change of variables $k \rightarrow p = k\dt$, however, this is no longer useful in our present situation.

For large $m\dt$, Kibble-Zurek physics indicates that once we are in the vicinity of the critical point, the only scale in the problem is the Kibble-Zurek time, $\tKZ = \sqrt{\dt/m}$. Then, it is promising to define dimensionless variables in this case as
\begin{eqnarray}
q & = &  k \, \tKZ=k \sqrt{\frac{\dt}{m}}  \,, \label{qqqq}\\
\kappa & = & = m \, \tKZ = \sqrt{m \dt}  \,.
\label{3-12}
\end{eqnarray}
Indeed this is the correct change of variables in the integral which allows us to extract the large $\kappa$ behaviour.

The renormalized expectation value becomes
\begin{eqnarray}
\sigma_s \vev{\phi^2}_{ren}|_{t=0} = \left(\frac{m}{\dt}\right)^{\frac{d-2}{2}} \int dq \left( \frac{q^{d-2}}{4\pi\sqrt{q^2+\kappa^2}} \left| \frac{E_{3/2}}{E'_{1/2} E_{3/2} - E_{1/2} E'_{3/2}} \right|^2 - f_{ct} (q,\kappa) \right) \,. \nnn \\ \label{modes_zero}
\end{eqnarray}
We will now show that the integrand has an expansion in $\frac{1}{\kappa}$, with the leading term being $O(\kappa^0)$. Therefore the leading large $\kappa$ behaviour is given by the pre-factor, which is in fact the expected KZ scaling, \ie
$1/\tKZ^{d-2}$, for this expectation value. 

At this point, it will be useful to remind the reader what the different $E$'s are in eq.~\reef{modes_zero}. In terms of the dimensionless variables introduced in eqs.~\reef{qqqq} and \reef{3-12}, we have
\begin{eqnarray}
E_{c} = \frac{\Gamma(c) \Gamma(b-a)}{\Gamma(b) \Gamma(c - a)} \  & , &  \ E'_{c} = E_{c} (a \leftrightarrow b)\,,
\nonumber \\
a = \alpha +\frac{i \kappa}{2}   \sqrt{\kappa ^2+q^2} & , & \ \ \ b  =  \alpha -\frac{i \kappa}{2} \sqrt{\kappa ^2+q^2}, \label{a and b 2} \\
\alpha  & = &   \frac{1}{4} \left(1-\sqrt{1-4 \kappa ^4}\right)  \,.  \nonumber
\label{3-13}
\end{eqnarray}
The usefulness of choosing the dimensionless momenta as in eq.~(\ref{3-12}) is the following: The crucial point is that in eq.~(\ref{a and b 2}) the expansion of $a$ is
\ben
a = \frac{1}{4}(1+iq^2)+\frac{i}{16\kappa^2} + \cdots \,,
\een
so that the leading term is $O(\kappa^0)$. This allows us to perform a series expansion of the integrand in inverse powers of $1/\kappa$. 
Note that the gamma functions which appear have vanishing arguments and therefore individually each $E$ can diverge. However, the combination present in eq.~(\ref{modes_zero}) is well-behaved. In fact, one gets an expansion
\begin{eqnarray}
\left| \frac{E_{3/2}}{E'_{1/2} E_{3/2} - E_{1/2} E'_{3/2}} \right|^2 =  \frac{\kappa}{8 \pi ^2}\,   e^{-\frac{5 \pi}{4}q^2} \left( e^{\pi  q^2}+1\right)^2 \left| \Gamma \left(\frac{1-i q^2}{4}\right) \Gamma \left(\frac{1+ i q^2}{2}\right) \right|^2  + O(1/\kappa) \,.  \nnn \\
\label{3-14}
\end{eqnarray}
With regards to the counterterm contributions \reef{ict} at $t=0$, all the terms that are proportional to the mass vanish and for lower dimensions, we do not have any time derivatives in the counterterms. Hence we need only consider the leading contribution proportional to $q^{d-3}$ from eq.~\reef{ict}.

As an aside, we note that the rescaling of the integration variable is simply a tool to obtain the large $\kappa$ behaviour. In fact, we can change to any other dimensionless variables, \eg
\begin{eqnarray}
\tilde{q} & = & k \frac{\dt^\beta}{m^{1-\beta}} \,, \\
\tilde{\kappa} & = & (m \dt)^\beta\,,
\end{eqnarray}
with $\beta$ being some real number. In the cases analyzed so far, $\beta=1$ for the fast quench and $\beta=1/2$ for the slow quench. For general $\beta$, one would obtain 
\begin{eqnarray}
a=\frac{1}{4} \left(-\sqrt{1-4 \kappa ^{2/\beta }}+\kappa ^{1/\beta } \left(2 i+\frac{i q^2}{\kappa^2}+O\left({1}/{\kappa^4}\right)\,\right)\right) \,.
\end{eqnarray}
Consider the term proportional to $q^2$. It turns out that if we chose $\beta <1/2$, then that term would be leading in the $\kappa$ expansion and $a$ would be just proportional to $q^2$. In the opposite case, with $\beta>1/2$, then that term would be subleading and $a$ won't depend on $q$ to leading order. It turns out that none of these possibilities allow us to get a well-behaved series expansion of the combination of $E$'s that we have in eq.~\reef{modes_zero}. It is only when $\beta=1/2$, that the leading term is at the same time independent of $\kappa$ {\it and} dependent on $q$ and that is exactly the right combination needed to produce an expansion in (inverse) powers of $\kappa^2$. Note that the natural scale in the problem is $\kappa$ and one might have thought that the expansion is in inverse powers of $\kappa$. However, the expansion in (\ref{3-14}) is in powers of $1/\kappa^2$, i.e. in powers of $ 1/\dt$.

Returning to the evaluation of $\vev{\phi^2}_{ren}$ at $t=0$, we observe that the leading term in eq.~(\ref{3-14}) is proportional to $\kappa$. Going back to eq. (\ref{modes_zero}), the factor $\frac{1}{\sqrt{q^2+\kappa^2}}$ starts with $1/\kappa$, and hence one finally gets for $d \leq5$,
\begin{eqnarray}
\sigma_s \vev{\phi^2}_{ren}|_{t=0} = \left(\frac{m}{\dt}\right)^{\frac{d-2}{2}} \int dq \, \Big[ \Phi_1(q) - q^{d-3} + \cdots \Big]\,, 
\label{3-15}
\end{eqnarray}
where $q^{d-3}$ corresponds to the counterterm contribution from eq.~\reef{ict} and
\begin{equation}
\Phi_1(q) =\frac{q^{d-2}}{8 \pi ^2}\,   e^{-\frac{5 \pi}{4}q^2} \left( e^{\pi  q^2}+1\right)^2 \left| \Gamma \left(\frac{1-i q^2}{4}\right) \Gamma \left(\frac{1+ i q^2}{2}\right) \right|^2  \,.
\label{phi1ofq}
\end{equation}
Now the integral over $q$ yields a numerical constant, so that the scaling behaviour is given by the prefactor, which is exactly the expected Kibble-Zurek scaling given in eq.~\reef{KZscalingCCP}.

We have not been able to perform the integral in eq. (\ref{3-15}) analytically.
We can, instead, integrate numerically to any desired precision. For example, in $d=4$ and $d=5$, we obtain
\begin{eqnarray}
d=4\ :&&\quad\sigma_s \vev{\phi^2}_{ren}|_{t=0}  =  \left(\frac{m}{\dt}\right) 0.091412 + \cdots \, ,\\
d=5\ :&&\quad\sigma_s \vev{\phi^2}_{ren}|_{t=0}  =  \left(\frac{m}{\dt}\right)^{3/2} 0.256921 + \cdots \,. \label{an_d5}
\end{eqnarray}
The expression obtained for $d=4$ fits perfectly with the purple curve in figure \ref{fig_kz_zero_sca}a. One can do an analogous calculation for $d=5$ and will find the same agreement, validating this analytic expansion. 

%Our analytic answer gives the corrections to the leading KZ scaling. A significant aspect of our result is that $\vev{\phi^2 }_{ren}$ has an expansion in inverse powers of $\kappa$ that is, in powers of $\dt^{-1/2}$. As we pass from the adiabatic regime to the critical regime, a power series expansion in $(m\dt)^{-1}$ breaks down, but a new expansion in inverse fractional powers of $(m\dt)$ is valid. This is similar to what was found in holographic studies of Kibble-Zurek scaling \cite{holo-kz}.

In higher dimensions, \ie $d\ge 6$, the counterterm contributions \reef{ict} also include terms involving time derivatives of the mass \cite{dgm1,dgm2}.  This means that even at $t=0$ the lower order counterterms can make a non-vanishing contribution. In fact, for $d=6$ and $7$, eq.~\reef{ict} yields
\begin{eqnarray}
f_{ct} (q,\kappa) = q^{d-3} + \frac{1}{4} q^{d-7} \,.
\end{eqnarray}
The last term will introduce the extra logarithmic divergence in $d=6$. In this case, we obtain,
\begin{eqnarray}
d=6\ :&&\quad\sigma_s \vev{\phi^2}_{ren}|_{t=0} =\left(\frac{m}{\dt}\right)^2 \left( \frac{1}{8}\,\log \!\big(\mu \,\dt\big) +0.030079 \right) +\cdots \,,
\end{eqnarray}
where we introduced a new renormalization scale $\mu$ in the logarithm, as in \cite{dgm1,dgm2,dgm3} for even $d$. Of course, this result perfectly matches the purple curve on figure \ref{fig_kz_zero_sca}b.

Extracting KZ scaling and corrections became much easier since we worked at $t=0$. In principle, one should be able to carry out the analysis for finite $t/\tkz$ in the critical region. We show how to deal with it analytically at the end of the next subsection. 

\subsection{Universality at any rate!}
\label{anyrate}

Above we focused our attention on $t=0$ where the TCP fermionic quenches \reef{fermionquench} and the CCP scalar quenches \reef{scalarquenchX} precisely reach the critical point. We found that the expectation values scale exactly as predicted by the Kibble-Zurek arguments. However, KZ scaling should hold not only at the critical point but also in its vicinity, see eq.~(\ref{t_over_tkz}), so in this section we study what happens with the quenched operators at {\it any} finite time. This analysis is also interesting because it will give a complete description of the expectation value of the quenched operator at any finite time for {\it any quench rate}.

Following our earlier studies \cite{dgm1,dgm2,dgm3}, we work in terms of the dimensionless time $\tau=t/\dt$.
The idea is first, to fix a finite value of $\tau = t/\dt = \tau_0$, and the study the response at this time as a function of $\dt$. 
This means that at different values of $\dt$, we are examining the response at different physical times $t = \tau_0\, \dt$.
In particular, recall that for the Kibble-Zurek time is given by $\tKZ = \sqrt{\dt / m}$ and hence we reach this time when
$\dt = \dt_\mt{KZ}=\tKZ/\tau_0 = 1/(m \tau_0^2)$.

As a function of $\dt$,  three different regimes will appear:
First, for $m \dt \ll 1$ we have the universal fast quench regime studied in \cite{dgm1,dgm2,dgm3}. At the other extreme, 
when $\dt \gg \dt_\mt{KZ}$, which means $m \dt \gg 1/\tau_0^2$, the system is far from the critical point and  the time evolution should be adiabatic. Finally for  $ 1 \ll m\dt \ll 1/\tau_0^2$, we will be in the KZ interval $|t|\lesssim\tkz$ and so we will observe Kibble-Zurek scaling. In general, the choice of $\tau_0$ is arbitrary but as a practical matter in the following examples, $\tau_0$ must be small enough (in absolute value) so that $ 1 \ll m\dt \ll 1/\tau_0^2$ is a large interval to make sure that the KZ scaling easily discernable. Note that $\dt_\mt{KZ}$ is inversely proportional to $\tau_0^2$, so as $\tau_0\to0$ the division between the KZ and adiabatic regimes diverges. This explains why we only saw the fast quench and KZ scaling regimes in the previous section where implicitly we set $\tau_0=0$.

\begin{figure}[h]
\setlength{\abovecaptionskip}{0 pt}
\centering
\includegraphics[scale=0.6]{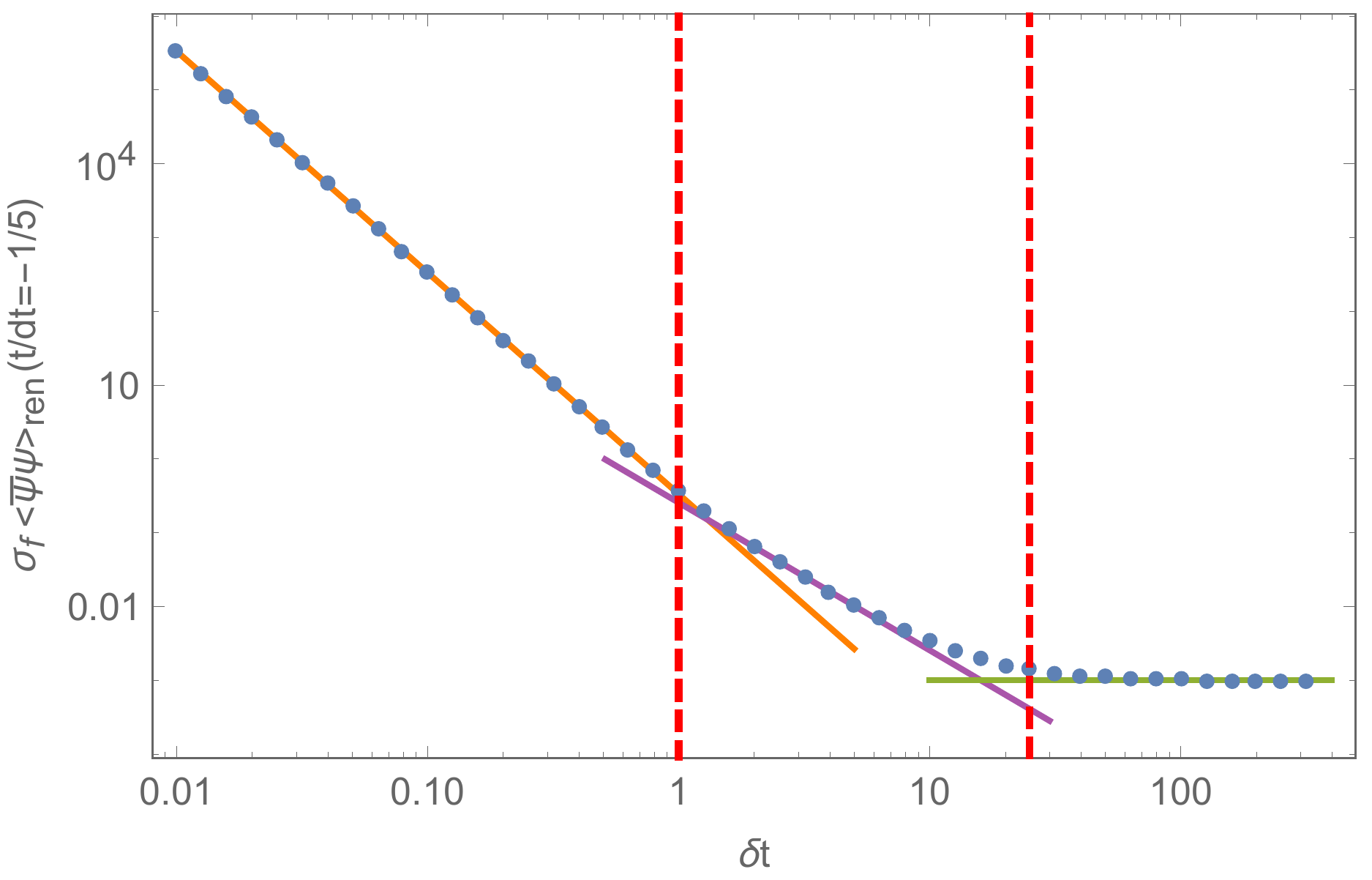}
\caption{Expectation value at fixed $\tau_0=-1/5$ as a function of $\dt$ for a TCP fermionic quench with $d=5$ and $m=1$. The solid orange line is the analytic leading contribution \reef{phi_odd} for fast quenches; the solid purple line is a linear best fit and agrees with the KZ scaling, $\vev{\bar{\psi}\psi}_{ren} \sim \dt^{-2}$; and the solid green line shows the adiabatic value for a fixed mass. As a guide to the eye, the dashed red lines show $\dt=1$ and $\dt=1/\tau_0^2$, which correspond to the transition regions. } \label{fig_fd5_tau}
\end{figure}

To illustrate the above discussion, we start by analyzing the TCP fermionic quench in $d=5$ with $m=1$ and $\tau_0=-1/5$. 
The expectation value of the mass operator for a wide range of $\dt$ is shown in figure \ref{fig_fd5_tau}. We can clearly recognize the three different scaling behaviours in this figure. First we have the fast quench scaling, whose analytic answer \reef{phi_odd} is plotted in solid orange. As $m\dt \to 1$, there is a transition and KZ scaling begins to appear. The solid purple line shows the expected KZ scaling \reef{KZscalingTCP}. Finally, when $\dt$ is large compared to $1/\tau_0^2$, the response becomes adiabatic and is independent of $\dt$.  The solid green line shows the value of the expectation value for a fixed mass \reef{ferm_adiab} with mass equal to $m(t/\dt = \tau_0)$. The passage between these three scaling behaviours appears to be completely smooth. 

\begin{figure}[h]
\setlength{\abovecaptionskip}{0 pt}
\centering
\includegraphics[scale=0.6]{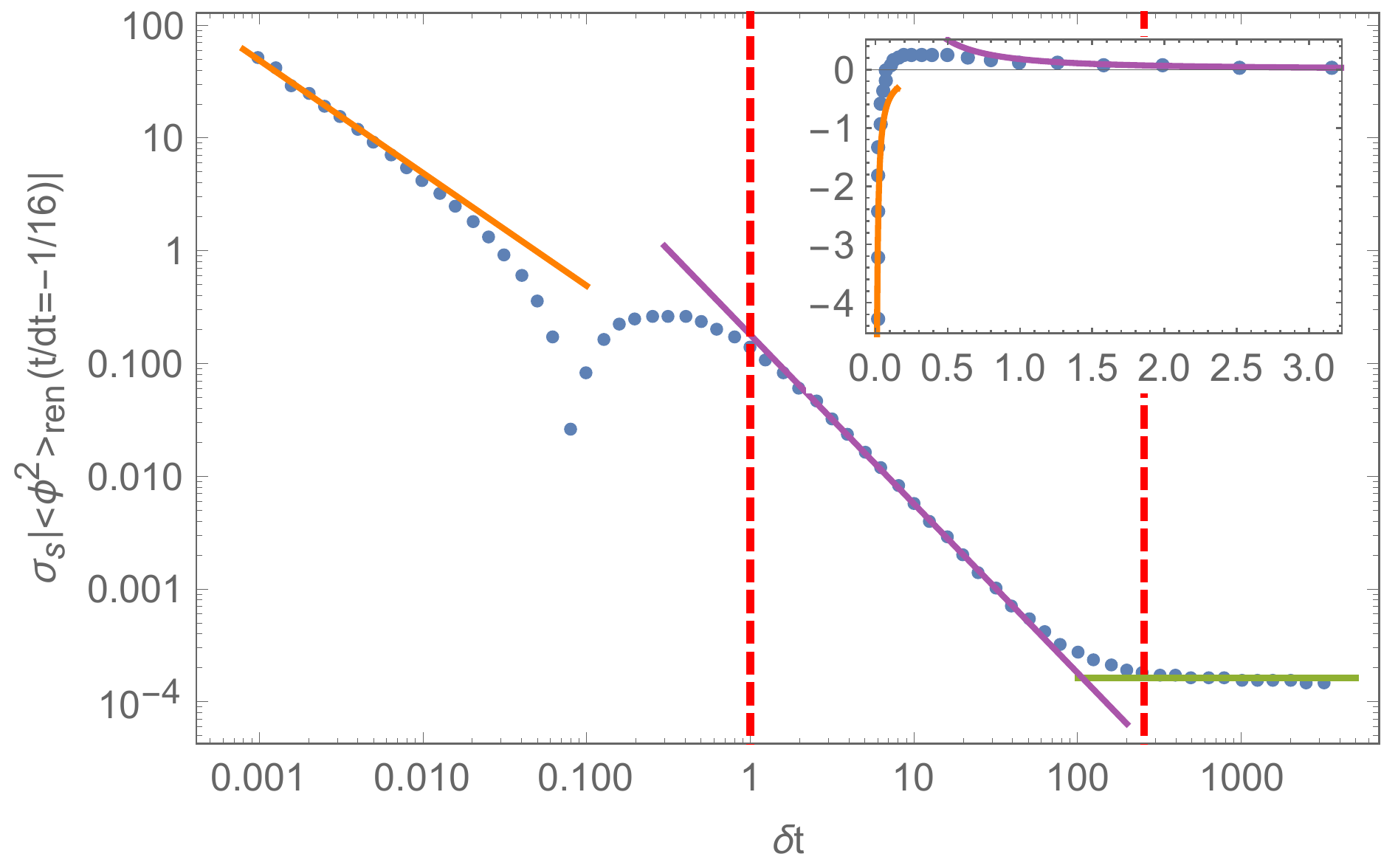}
\caption{Expectation value at fixed $\tau=-1/16$ as a function of $\dt$ for a CCP scalar quench with $d=5$ and $m=1$. The solid orange line is the analytic leading contribution \reef{boat} for fast quenches; the solid purple line is a linear best fit and agrees with the KZ scaling, $\vev{\phi^2}_{ren} \sim \dt^{-\frac{3}{2}}$; 
%(the fit by a function $y= a x^{-\alpha}$ gives $a=0.1867$ and $\alpha=1.515$) ; 
and the solid green line shows the adiabatic value for a fixed mass. As a guide to the eye, the dashed red lines show $\dt=1$ and $\dt=1/\tau_0^2$, which correspond to the transition regions. Note that the expectation value changes sign near $\dt\sim 0.1$ which produces a dramatic effect on the logarithmic scale. To avoid misinterpretations, we included an inset showing that the expectation value is a smooth function of $\dt$ at this point.} \label{fig_sd5_tau}
\end{figure}

The next example is a CCP quench for the scalar field. Figure \ref{fig_sd5_tau} shows the expectation value for $d=5$, $m=1$ and $\tau_0=-1/16$. Note that since we are now away from $t=0$, we can also see the fast quench scaling in odd dimensions. The results are essentially the same as in the previous example of a fermionic quench. There are three distinct phases for the scaling of the expectation value  as a function of $\dt$. For small $\dt$, we see the fast quench scaling with $\vev{\phi^2}_{ren}\sim1/\dt$. For very large $\dt$, the expectation value is just the adiabatic one, independent of $\dt$. But between these two regimes, there is a Kibble-Zurek scaling in the region $1 < m\dt < 1/(\tau_0^2)$, where  $\vev{\phi^2}_{ren}\sim1/\dt^{{3}/{2}}$. As observed in the previous section, the expectation value smoothly changes sign between the fast quench and KZ regimes, which, however, produces a rather dramatic effect on the logarithmic scale.

\subsection{Scaling functions} \labell{scalingfns}

We can understand the KZ behaviour analytically in the case of the CCP scalar quench with a pulsed mass profile \reef{scalarquenchX} by generalizing the arguments in section \ref{ana_pulsed} to a finite fixed (dimensionless) time $t/\dt$. In particular, by making an appropriate expansion of the full expectation value, we will find that takes exactly the form claimed in eq.~(\ref{t_over_tkz}), \ie an overall scaling factor times a function of $t/\tKZ$.

Recall that the (bare) expectation value for the pulsed scalar quench is given by
\begin{eqnarray}
\vev{\phi^2 (t/\dt) } = \int \frac{d^{d-1}k}{(2\pi)^{d-1}} |u_\vk (t/\dt)|^2 \,,
\end{eqnarray}
where the mode solutions are given by
\begin{eqnarray}
u_\vk & = & \frac{1}{\sqrt{4\pi}(k^2+m_0^2)^{1/4}} \frac{2^{i \sqrt{k^2+m_0^2}} y^\alpha}{E'_{1/2} E_{3/2} - E_{1/2} E'_{3/2}}  \label{modes_an} \\ 
& & \times \left( E_{3/2} \ _2F_1 (a,b;\frac{1}{2};1-y) + E_{1/2} \sinh(t/\dt) _2F_1 (a+\frac{1}{2},b+\frac{1}{2};\frac{3}{2};1-y) \right). \nonumber 
\end{eqnarray}
with $y=\cosh^2(t/\dt)$. The latter was used in section \ref{ana_pulsed} to simplify the expression as $1-y|_{t=0}=0$ and then both hypergeometric functions in eq.~(\ref{modes_an}) had a vanishing argument and so they simplified to 1. 

Here rather than setting $t=0$, we want to expand for small $t/\dt$. Note that according to the arguments presented in section \ref{kzphysics}, we expect the expectation value to take the form in eq.~(\ref{t_over_tkz}) for $|t|\lesssim\tkz$. Then it will be useful to write the fixed (dimensionless) time $t/\dt$ as
\begin{eqnarray}
\frac{t}{\dt} = \frac{t/\tKZ}{\kappa} \,,
\label{tkz}
\end{eqnarray}
where again $\tkz = \sqrt{\dt/m}$ as in eq.~\reef{ttkz} and $\kappa = \sqrt{m \dt}$ is the dimensionless mass introduced in eq.~\reef{3-12}. As in eq.~\reef{qqqq}, we also scale the momentum by $\tkz$ to define $q = k \sqrt{\frac{\dt}{m}}$. In terms of these dimensionless variables, the expectation value becomes,
\begin{eqnarray}
\sigma_s \vev{\phi^2 (t/\tkz) } =  & & \left(\frac{m}{\dt}\right)^{\frac{d-2}{2}} \int dq \frac{q^{d-2}}{\sqrt{q^2+\k^2}} \left|\frac{E_{3/2}}{E'_{1/2} E_{3/2} - E_{1/2} E'_{3/2}}\right|^2   \labell{golly}\\
& & \times \left| _2F_1 (a,b;\frac{1}{2};1-y)+ \frac{E_{1/2}}{E_{3/2}} \sinh \left(\frac{t/{\tkz}}{\kappa}\right) {}_2F_1 (a+\frac{1}{2},b+\frac{1}{2};\frac{3}{2};1-y) \right|^2 \nnn
\end{eqnarray}
Now from eq.~\reef{tkz} for times of order $t/\tKZ \lesssim 1$ where we expect to observe the KZ scaling, we see that examining small $t/\dt$ is equivalent to studying the limit of large $\kappa$.  Hence our approach will be to evaluate the integral in eq.~\reef{golly} in this large $\kappa$ limit and the leading order contribution should result in a scaling function  $F(t/\tkz)$.

First we note that the factors appearing in the first line of the integrand in eq.~\reef{golly} are precisely the $t=0$ integrand analyzed in Section \ref{ana_pulsed} --- see eq.~\reef{phi1ofq}. Hence we already have the large $\kappa$ limit of these terms being 
\beqa
&&\frac{q^{d-2}}{\sqrt{q^2+\k^2}} \left|\frac{E_{3/2}}{E'_{1/2} E_{3/2} - E_{1/2} E'_{3/2}}\right|^2 \xrightarrow[\kappa \to \infty]{} \labell{wacko}\\
&&\qquad\qquad \Phi_1(q) \equiv \frac{ q^{d-2}}{8 \pi ^2}\,  e^{-\frac{5 \pi}{4}q^2} \left( e^{\pi  q^2}+1\right)^2\, \left| \Gamma \left(\frac{1-i q^2}{4}\right) \Gamma \left(\frac{1+ i q^2}{2}\right)\right|^2   \,.\nnn
\eeqa
Next, we need to expand $\frac{E_{1/2}}{E_{3/2}} \sinh \left(\!\frac{t/\tkz}{\kappa}\!\right)$ for large $\k$. In this limit, the leading term from $\sinh$ is proportional to $1/\k$ but expanding the ratio  ${E_{1/2}}/{E_{3/2}}$ gives a leading term proportional to $\k$. Combining these, we have
\beq
\frac{E_{1/2}}{E_{3/2}} \sinh \left(\frac{t/\tkz}{\kappa}\right) \xrightarrow[\kappa \to \infty]{} \frac{t}{\tkz}\,  \frac{2\, e^{3\pi i/4} \, \Gamma \left(\frac{3-i q^2}{4}\right)}{\Gamma \left(\frac{1-i q^2}{4}\right)} + O(\k^{-2}) \,.
\eeq

The last step is to expand the hypergeometric functions in eq.~(\ref{modes_an}) to leading order for large $\kappa$. Note that by expanding its arguments we get,
\begin{eqnarray}
\lim_{\kappa \to \infty} {}_2F_1 (a,b;c,1-y) = \lim_{\kappa \to \infty} {}_2F_1 \left(a, -i \kappa^2;c,-\frac{t/\tkz}{\kappa^2}\right) = \lim_{\kappa \to \infty} {}_2F_1 \left(a, \kappa^2;c, i \frac{t/\tkz}{\kappa^2}\right) \,, \nnn 
\end{eqnarray}
where $a= \frac{1+i q^2}{4} + O(\k^{-2})$ and $c=1/2$ or $3/2$ depending which of the two hypergeometric functions we are considering. To produced the second equality, we used the series representation of the hypergeometric function: ${}_2F_1 (a,b;c,z)=\sum\frac{(a)_n (b)_n}{(c)_n} \frac{z^n}{n!}$. With $b=-i\k^2$, we have $(b)_n\simeq (-i \k^2)^n$ to leading order\footnote{Recall $(X)_n \equiv X(X+1) \cdots (X+n-1)$ and $(X)_0=1$.} and so the $(-i)^n$ can be transferred to the last factor as $(-iz)^n$. Note also that the second hypergeometric in eq. (\ref{modes_an}) has $b+1/2$ as the second term, however, that extra $1/2$ will be irrelevant in the large $\kappa$ limit.
Now the limit $\k \to \infty$ yields confluent hypergeometric functions with the identity 
\beq
\lim_{w\to \infty} {}_2F_1 (x,w; y;z/w) = {}_1F_1 (x; y;z) = \sum_{n=0}^\infty \frac{(x)_n}{(y)_n}\, \frac{z^n}{n!}\,.
\labell{identity}
\eeq
Hence the hypergeometric functions in eq.~(\ref{modes_an}) become
\begin{eqnarray}
_2F_1 (a,b;1/2;1-y) & \xrightarrow[\kappa \to \infty]{} & {}_1F_1 \left(\frac{1+i q^2}{4}, \frac12,\frac{ i\, t^2}{\tkz^2} \right) +O(\kappa^{-2}) \,, \\
_2F_1 (a+1/2,b+1/2;3/2;1-y) & \xrightarrow[\kappa \to \infty]{} & {}_1F_1 \left(\frac{3+i q^2}{4}, \frac32,\frac{ i\, t^2}{\tkz^2} \right) +O(\kappa^{-2}) \,,
\end{eqnarray}
both of which are independent of $\k$ to leading order.

With this we have all the ingredients to compute the bare expectation value. The final component is the counterterm contributions \reef{ict} needed to regulate the expectation value. For $d\le 5$, the necessary contributions can be written as
\begin{eqnarray}
f_{ct} (q,\k) = q^{d-3} - q^{d-5} \, \frac{\k^2}{2} \tanh^{2} \!\left(\frac{t/\tkz}{\kappa} \right) = q^{d-3} - q^{d-5} \, \frac{t^2}{2\,\tkz^2} + O(\k^{-2}) \,. 
\end{eqnarray}
 
\begin{figure}[H]
        \centering
        \subfigure[Long times, $-4<t/\tkz<4$.]{
                \includegraphics[scale=0.39]{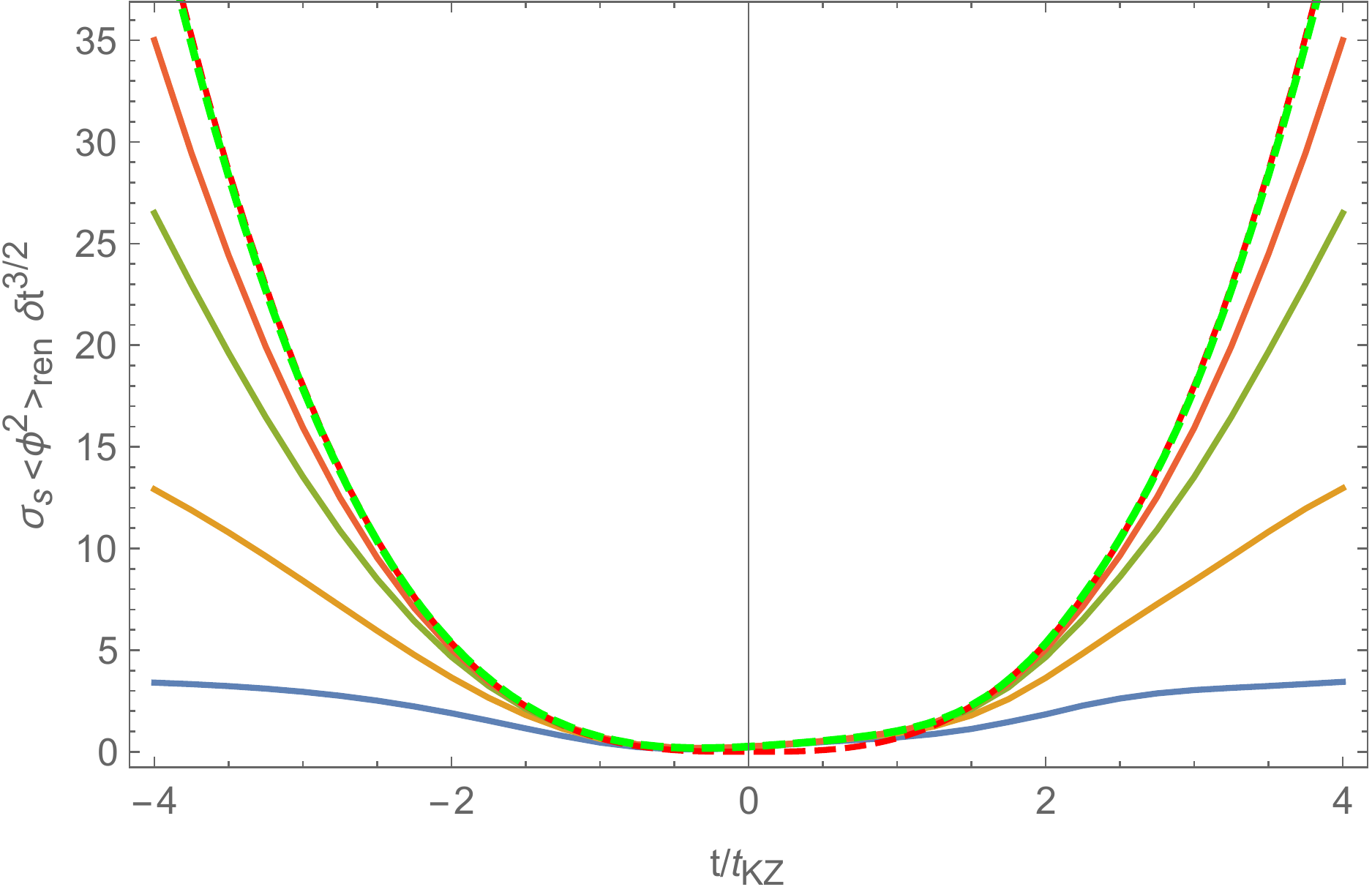} }
   		 \subfigure[Kibble-Zurek interval, $-1<t/\tkz<1$.]{
                \includegraphics[scale=0.39]{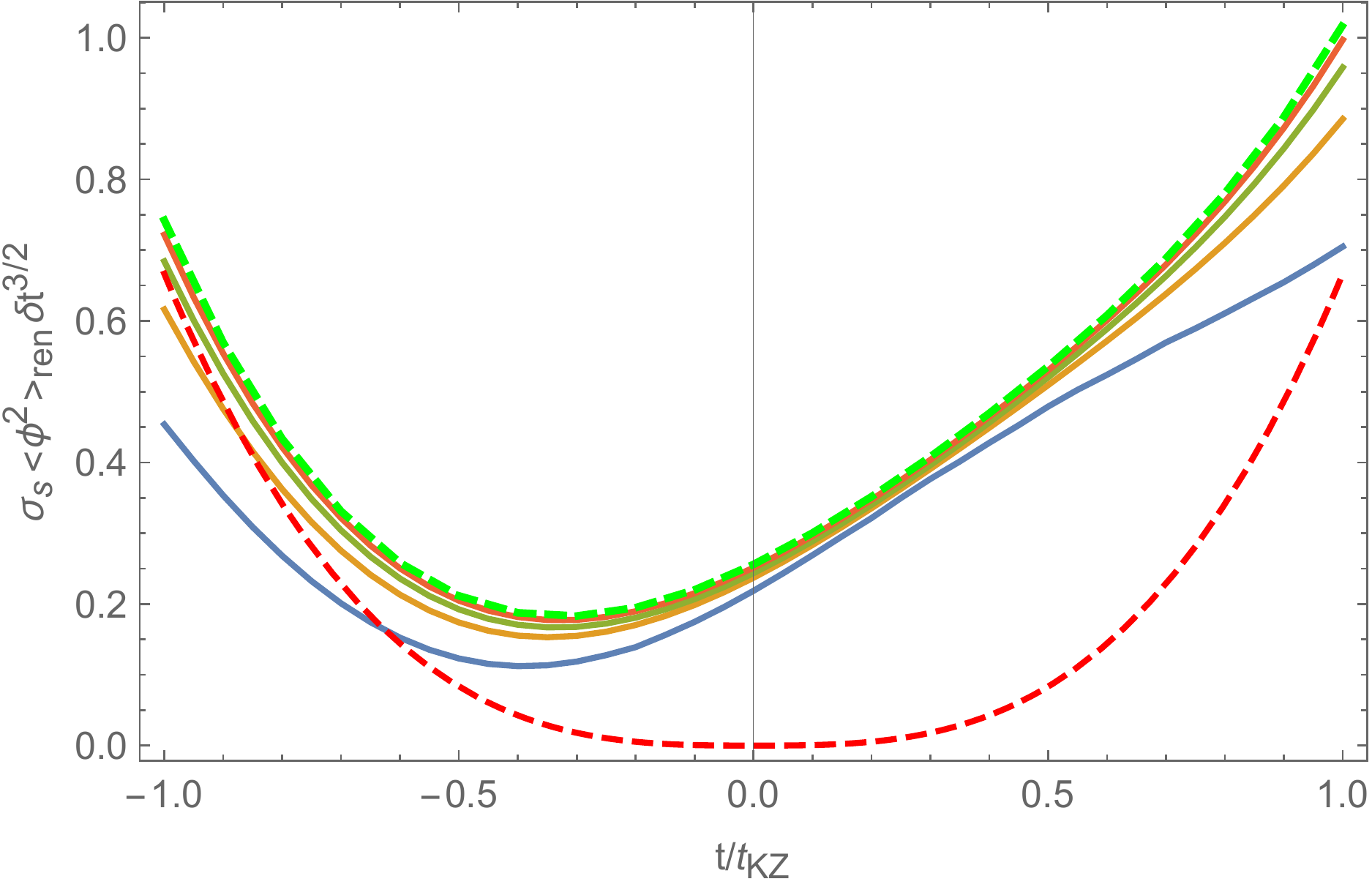} }
        \caption{Renormalizad expectation value of $\phi^2$ as a function of $t/\tkz$ for $d=5$. The different curves correspond to $m\dt=10^{i}$, with $i=0.5 \text{(blue)},\ 1 \text{(yellow)},\ 1.5 \text{(green)},\ 2 \text{(orange)}$. Note that we are multiplying the expectation value by $(\dt/m)^{\frac{3}{2}}$ to cancel the expected KZ scaling. In dashed red, we show the adiabatic result for the expectation value. In dashed green, we present the scaling solution emerging from the large $\k$ expansion. Panel (a) shows the results over longer time periods, while panel (b) zooms in the interval $-1<t/\tkz<1$.
\label{fig_evidence_sca_2}}
\end{figure}

Then the renormalized expectation value to leading order for large $\kappa$ is given by
\ben
\sigma_s \vev{\phi^2 (t/\tkz)}_{ren} = \left(\frac{m}{\dt}\right)^{\frac{d-2}{2}} F(t/\tkz)
\labell{soap}
\een
with
\begin{eqnarray}
& & F(t/\tkz)=  \int dq \, \bigg[ \Phi_1(q)  
\left| {}_1F_1 \left(\frac{1+i q^2}{4}, \frac{1}{2}, \frac{ i\, t^2}{\tkz^2} \right) + \frac{t}{\tkz} \,\frac{2\, e^{3\pi i/4} \,\Gamma \left(\frac{3-i q^2}{4}\right)}{\Gamma \left(\frac{1-i q^2}{4}\right)} \, {}_1F_1 \left(\frac{3+i q^2}{4}, \frac{3}{2}, \frac{ i\, t^2}{\tkz^2} \right) \right|^2 \nnn \\
& &\qquad\qquad\qquad \qquad\qquad\qquad\qquad\qquad\qquad  -\,  q^{d-3} + q^{d-5} \, \frac{t^2}{2\,\tkz^2} \,  \bigg] \,. \label{analytic}
\end{eqnarray}
Of course, this is just what we were looking for! The overall factor in eq.~\reef{soap} gives the correct Kibble-Zurek scaling and the expression in eq.~\reef{analytic} is an integral over $q$ that only depends on $t/\tkz$. Hence performing the integral yields a scaling function $F(t/\tkz)$, as expected in eq. (\ref{t_over_tkz}). In fact, the integral can be done numerically and so, we can compare with the full expectation value for different values of $\dt$. Basically, in figure \ref{fig_evidence_sca_2}, we present the analogue for CCP scalar quenches of figure \ref{fig_evidence_fer_2} for the TCP fermion quenches but with the added green dashed curve that gives the scaling function, \ie the leading order solution in the $\kappa$ expansion, computed by numerically integrating eq.~(\ref{analytic}). We observe that as $\dt$ increases, the full solutions approach $F(t/\tkz)$  in the range $|t|\lesssim \tkz$ and clearly move away from the adiabatic result, providing good evidence of the expected Kibble-Zurek scaling.

It is also worth mentioning that this analytical computation agrees with the numerical fit in the previous section. For instance, for the CCP scalar quench illustrated in figure \ref{fig_sd5_tau}, the KZ scaling regime was fit with the purple curve as $y= a \dt^{-\alpha}$, with $a=0.1867$ and $\alpha=1.515$. The expected value of the exponent for the KZ scaling in $d=5$ is $\alpha=3/2$ and so the fit gives good agreement with this. But we also find quite a good agreement in the overall coefficient $a$ here. In figure \ref{fig_sd5_tau}, $\tau_0$ is fixed to $-1/16$ and $\dt$ is of the order of 10 in the KZ region, so $t/\tkz \sim - 0.2$. The numerical integration of eq.~(\ref{analytic}) for $d=5$ gives $a \sim 0.194$, which is again close to the numerical fit above. 

Figure \ref{fig_evidence_sca_2} also reveals another interesting feature about the scaling function $F(t/\tkz)$. One's initial impression might be that this function is {\it only} appropriate to describe the Kibble-Zurek region, \ie $|t|\lesssim \tkz$. However, figure \ref{fig_evidence_sca_2}a shows that for $|t|>\tkz$, $F(t/\tkz)$ overlaps with the adiabatic curve, showing that it also describes the behaviour of the expectation value in the adiabatic regime. Even though this might be surprising, it also be related to the fact that to obtain $F(t/\tkz)$ we just performed an expansion for large $\kappa$ expansion but we did not assume any special limit for $t/\tkz$.

\begin{figure}[h]
\setlength{\abovecaptionskip}{0 pt}
\centering
\includegraphics[scale=0.35]{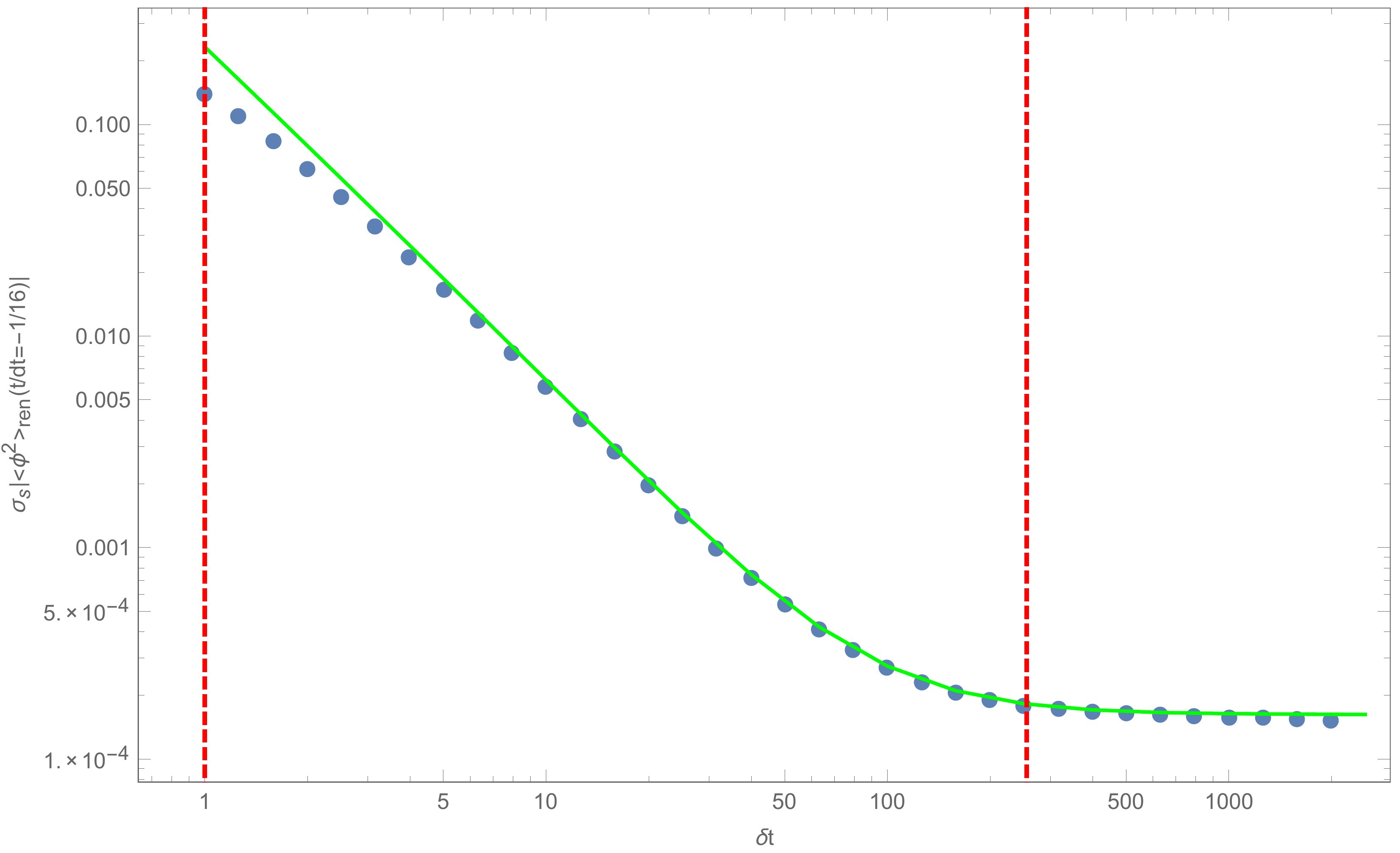}
\caption{Expectation value $\vev{\phi^2}_{ren}$ for the CCP scalar quenches at fixed $t/\dt=\tau_0=-1/16$, as a function of the {\textit{large}} $\dt$ with $d=5$ and $m=1$. The same numerical results are shown in figure \ref{fig_sd5_tau}. The solid green line shows the leading order solution in the $\k$ expansion, $\dt^{-3/2} F(-\sqrt{\dt}/16)$. As a guide to the eye, the dashed red lines indicate $\dt=1$ and $\dt=1/\tau_0^2$, which roughly separate the different scaling regimes. } \label{fig_sd5_F}
\end{figure}

Moreover, we can extend this discussion by comparing $F(t/\tkz)$ with the full numerical evaluation of figure \ref{fig_sd5_tau}. There, we fixed $t/\dt = -1/16$. Then, in order to compare both solutions we need to compute $F(t/\tkz)=F(\frac{t}{\dt} \sqrt{m \dt}) = F(-\frac{1}{16} \sqrt{\dt})$ with $m=1$. Figure \ref{fig_sd5_F} compares the numerical evaluation of the full solution with $F(-\frac{1}{16} \sqrt{\dt})$ for large $\dt$. The overlap between the two curves at large $\dt$ makes manifest that $F(t/\tkz)$ is a good approximation even during the adiabatic evolution. 

Further let us observe that the KZ scaling function \reef{analytic} was derived by taking a limit where $\kappa=\sqrt{m\dt}$ becomes large. Since $m\dt$ is the only dimensionless quantity characterizing the free field quenches, this should also correspond to the adiabatic limit, \ie the limit of large $\dt$.  This perspective suggests that it is natural to expect some kind of agreement between the two solutions. 
However, this agreement is still somewhat counterintuitive. Using eqs.~\reef{scalarquenchX} and \reef{sca_adiab}, the adiabatic expectation value for $d=5$ is expected to be
\beq
\sigma_s\,\vev{\phi^2(t/\dt)}_{ren}=\frac23\,m^3\,\tanh^3(t/\dt)\,,\labell{oneX}
\eeq
which for large $t/\dt$ saturates at the constant value, $\sigma_s\,\vev{\phi^2}_{ren}|_{t\to\infty}=\frac23\,m^3$. On the other hand, for $d=5$, the KZ scaling solution \reef{soap} takes the form
\beq
\sigma_s\,\vev{\phi^2(t/\tkz)}_{ren}=\left(\frac{m}\dt\right)^{3/2}\,F(t/\tkz)\qquad{\rm with}\ \ \tkz=\sqrt{\dt/m}\,,\labell{twoX}
\eeq
which simply can {\it not} saturate to a constant proportional to $m^3$ at large $t/\tkz$. To resolve this tension, we note the key difference between eqs.~\reef{oneX} and \reef{twoX} is the scale with which the time is compared. That is, eq.~\reef{oneX} implicitly holds $t/\dt$ fixed while eq.~\reef{twoX} holds $t/\tkz$ fixed. Holding the second ratio fixed in the adiabatic solution instead yields 
\beqa
\sigma_s\,\vev{\phi^2(t/\tkz)}_{ren}&=&\frac23\,m^3\,\tanh^3\left(\frac1\k\,\frac{t}\tkz\right)\nnn\\
&=&\frac23\,\left(\frac{m}\dt\right)^{3/2}\,\left(\frac{t}\tkz\right)^3\,\left(1-\frac1{\k^2}\,\frac{t^2}{\tkz^2}+\cdots\right)\,,\labell{oneY}
\eeqa
Hence we see that this particular limit will yield agreement with the KZ scaling solution if for large $t/\tkz$, the scaling function reduces to $F(t/\tkz)\simeq \frac23\,\left({t}/\tkz\right)^3$. 
%\comment{Can we get this analytically?? DAG: Not yet :(, but I'm still trying.} 
Of course, the agreement in figure \ref{fig_evidence_sca_2} shows that this must be the case.\footnote{It is straightforward to generalize this result to general $d$. In this case, the large $t/\tkz$ behaviour of the scaling function becomes
$\sigma_s\,\vev{\phi^2(t/\tkz)}_{ren}= \frac{\Gamma \left(1-\frac{d}{2}\right) \Gamma \left(\frac{d-1}{2}\right)}{2 \sqrt{\pi }}\,(t/\tkz)^{d-2}$.} To contrast with our initial discussion above, let us further note that in the regime where the adiabatic and KZ scaling solutions agree in this figure, both curves are rapidly increasing rather than approaching a constant value.  We have also kept the first correction in the large $\k$ expansion to this simple scaling behaviour in eq.~\reef{oneY}. This term makes clear that if we are working with large but finite $\kappa$, then we should expect the agreement between the adiabatic solution \reef{oneX} and the KZ scaling solution \reef{twoX} to break down when $t/\tkz\simeq\k$ --- see further discussion in section \ref{discuss}.

\section{Results for ECPs}
\label{ECP}

Finally we consider End-Critical Protocols in the free scalar field theory by examining the tanh quenches \reef{ECPX} which end at zero mass. We already studied the early time scaling in fast quench regime for these quenches in \cite{dgm1,dgm2,dgm3} and so in this section, we will concentrate on the late time scaling in the slow quench regime. In particular, we are interested in the appearance of Kibble-Zurek scaling as the mass approaches $m=0$. At late times, the mass profile in eq.~\reef{ECPX} decays exponentially with $m(t)=m\, \exp(-t/\dt)$. As emphasized in section \ref{kzphysics}, the description of the KZ behaviour is slightly different for this exponential approach to the critical point, in comparison to the power law approach of the TCPs and CCPs examined in the previous section. In particular, rather than focusing on the time at which adiabaticity breaks down, we consider the value of the gap $\ekz$ at this point and then the KZ scaling takes the simple form given in eq.~\reef{1-4X}. For the case at hand, $\ekz=1/\dt$ and this scaling becomes the result given in eq.~\reef{KZscalingECP}, \ie $\vev{\phi^2}_{ren}\simeq 1/\dt^{d-2}$.

Our approach here is similar to that in section \ref{anyrate}. That is, we fix the dimensionless ratio $t/\dt=\tau_0$ to a sufficiently large value and then evaluate the expectation value as a function of $\dt$. Let us note that with the exponential decay of the mass, adiabaticity always breaks down irrespective of the parameters. In particular, with $m(t)=m\, \exp(-t/\dt)$, eq.~\reef{1-1} is satisfied for $\tkz = \dt \,\log(m\dt)$. Hence, when we fix $t/\dt=\tau_0$, we will be in the KZ scaling regime for
\beq
1\ \lesssim\ m\dt\ \lesssim\ m\delta\tkz \equiv \exp\tau_0 \,.
\label{ecp_Eq}
\eeq
For larger values, \ie $\m\dt>m\delta\tkz$, the response would be adiabatic, and for smaller values, \ie $\m\dt<1$, we would be in the fast quench regime.
One practical issue, however, is that identifying the early time scaling for the fast quenches is very difficult here because we are examining large values of $t/\dt$ and $\vev{\phi^2}_{ren}$ decays very rapidly after the initial quench. This issue could be avoided by examining the energy density, which is conserved in a global quench. Of course, the fast quenches of this kind were already extensively studied in \cite{dgm1,dgm2,dgm3}. Hence, for simplicity and for cohesion with the rest of the paper, we continue examining the expectation value of the mass operator but focus only on the KZ scaling and adiabatic regimes in the following examples. 

\begin{figure}[h]
\setlength{\abovecaptionskip}{0 pt}
\centering
\includegraphics[scale=0.5]{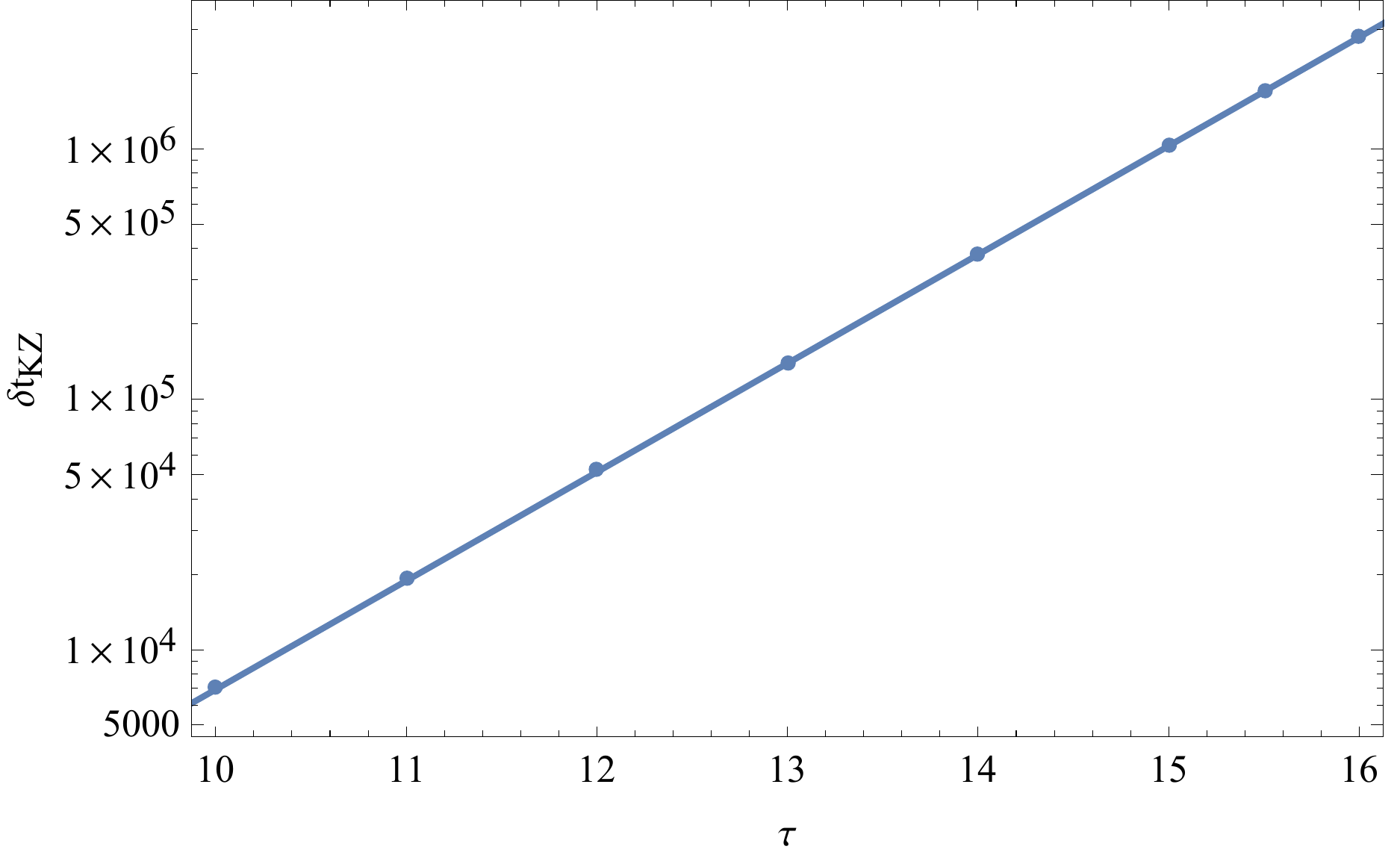}
\caption{The transition value of $\dt$  between the KZ scaling and the adiabatic behaviour as a function of the fixed $\tau_0=t/\dt$ (with $m=1$ and $d=5$). Note that the vertical axes is on a logarithmic scale. The best linear fit (in solid blue) with $\log(\delta\tkz) = a + b \, \tau_0$ yields $a=-1.0783$ and $b=0.9948$, supporting the exponential relation between $\delta\tkz$ and $\tau_0$.} \label{fig_ecp2}
\end{figure}

In figure \ref{fig_ecp2}, we show $\vev{\phi^2}_{ren}$ as a function of $\dt$ for $t/\dt=12$, $m=1$ and $d=5$. In the KZ scaling regime, we made a linear best fit which yields: $y= a \, \dt^{-\alpha}$, with $a=0.0199$ and $\alpha=2.993$. Hence, the fit agrees with the expected KZ scaling \reef{KZscalingECP} \ie $\vev{\phi^2}_{ren} \sim 1/\dt^{3}$ for $d=5$. We also see the expected transition to the adiabatic behaviour at roughly $\dt=\delta\tkz$.
\begin{figure}[h]
\setlength{\abovecaptionskip}{0 pt}
\centering
\includegraphics[scale=0.5]{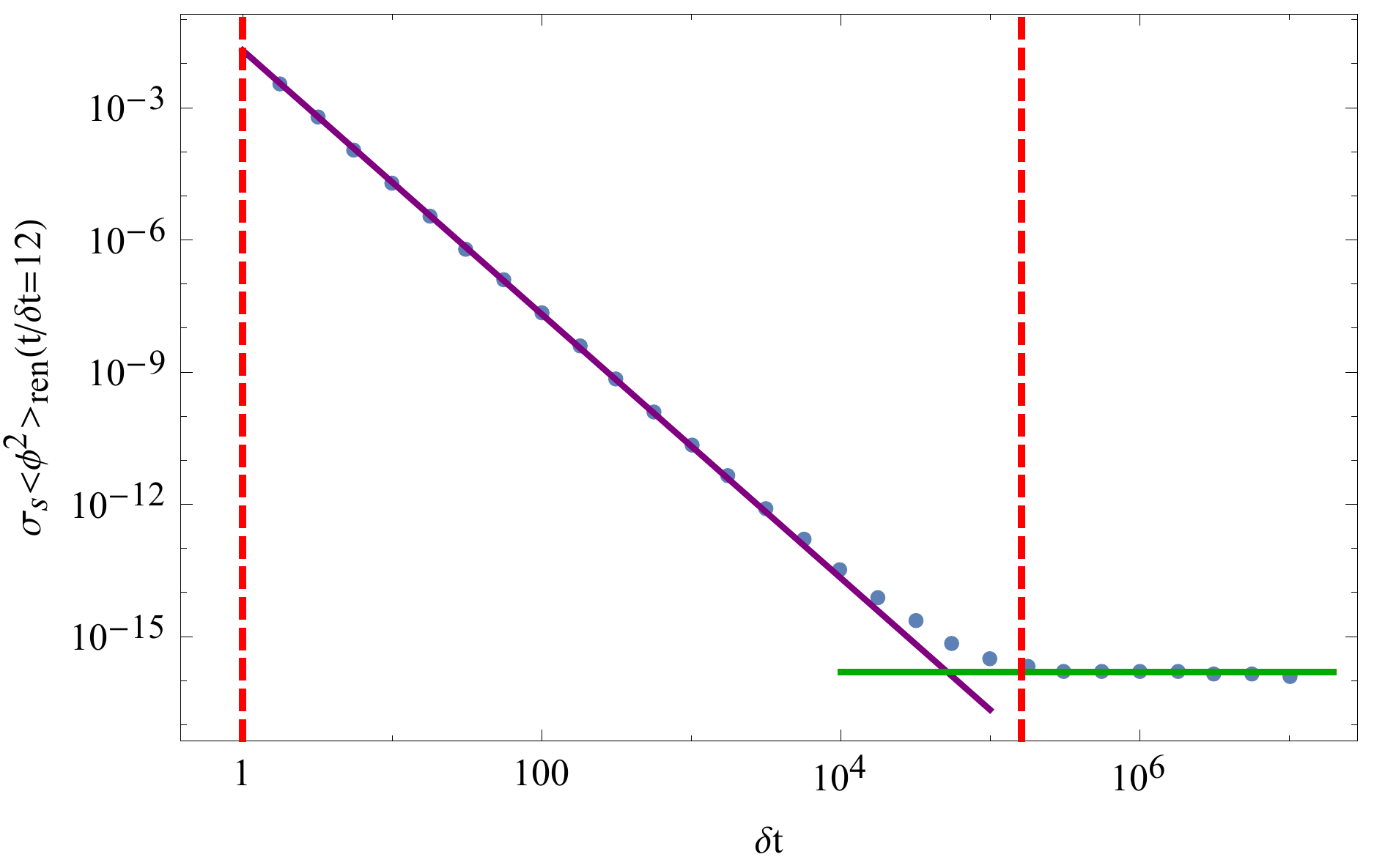}
\caption{Expectation value of the scalar mass operator as a function of $\dt$ with fixed $t/\dt=12$, and $d=5$, $m=1$. The solid purple line is the linear best fit for the points in the KZ regime. The green line corresponds to the adiabatic value at $t/\dt=12$. Red dashed lines correspond to $\dt=1$, the transition from the fast to the slow quench and $\dt=\dt_{KZ} = \exp(\tau_0)=\exp(12)$, the transition from Kibble-Zurek to adiabatic.} \label{fig_ecp1}
\end{figure} 

Figure \ref{fig_ecp2} shows the results of a second computation to the expression for $\delta\tkz$ in eq.~(\ref{ecp_Eq}). In particular, we estimated $\delta\tkz$ by computing the intersection between the linear fit for the KZ regime and the adiabatic value for different values of $t/\dt$. The results in the figure \ref{fig_ecp2} show that as expected $\delta\tkz$ grows exponentially with $\tau_0=t/\dt$. 

\section{Concluding Remarks} \labell{discuss}

In this paper we have studied various mass quenches in free field theories for a wide range of quench rates. We were able to exhibit universal scaling of the expectation value of the mass operators in both the fast and slow quench regimes. In particular, we found that the fast quench scaling smoothly crosses over to Kibble-Zurek scaling and finally to an adiabatic behaviour. 

\subsection*{Kibble-Zurek beyond free field theory}
Previously, in \cite{dgm1,dgm2,dgm3}, we showed that the fast scaling that was present in both free field and holographic theories should be valid in general interacting theories. In this work we show that KZ scaling is present in free field theory, so a natural question would be whether this scaling also holds beyond the present setup and in a much broader class of field theories. Again, as in the case of the fast quench, we have holographic studies that support the KZ scaling \cite{holo-kz,holo-kz2}, but having a more general argument indicating this scaling would certainly be an interesting direction for further study.

As pointed out in the Introduction, the KZ argument also involves predictions on the density of defects in the KZ phase. It might also be interesting to study models where we can actually compute defect formation and then extend the calculations to the fast quench regime, where we might also see a new novel scaling appearing for the defect density.

Finally, recently the appearance of KZ physics has been reported for systems in which the quench parameter is treated as a dynamical field \cite{Kolodrubetz:2014nia}. It would be very interesting to see what are the implications of our results in this new context.\footnote{A related model where the driving parameter is a dynamical field and becomes trapped near the critical point was studied in \cite{Kofman:2004yc}. We thank one of our anonymous referees for pointing out these references.}

\subsection*{UV cutoff and instantaneous quenches}

Our calculations involved {\em renormalized} expectation values of {\em local} operators and so as discussed in \cite{dgm1,dgm2,dgm3}, the quench rate is always much slower than the UV cutoff scale, \ie $\Lambda_\mt{UV}\gg1/\dt$. With these operators, it is not possible to study the case of {\em instantaneous} quenches, \ie $\dt \rightarrow 0$. However, as shown in some detail in \cite{dgm3}, certain properties of instantaneous quenches can be studied by looking at UV finite objects, such as  correlation functions at finite spatial separations.
In the latter case, the separation $r$ provides an extra scale in the problem somewhat analogous to the UV cutoff and we found that for $\dt<r$, the early time scaling \reef{0-4} saturated and the correlator became independent of $\dt$.

We expect a similar behaviour for local quantities when the UV cutoff is finite. Such models may describe realistic experimental systems and so it would be interesting to analyze those cases. In that case, one can consider a quench rate which is at the cutoff scale. For such rates, the physics should be described by an instantaneous quench. For such quenches, Calabrese and Cardy have proposed a simple description of the state after a quench from a gapped phase to a critical theory \cite{cc2,cc3} in terms of boundary states of the final CFT. It turns out that the validity of this proposal depends on which observable is being measured \cite{mandal}. An exactly solvable model on a lattice will be useful to address these issues. Progress in this direction has been made recently in \cite{Francuz:2015zva} analyzing the Kibble-Zurek scaling in the transverse Ising model.
We have recently found exactly solvable quench protocols in several spin models and studied the dependence on the quench rate: the results will appear in a separate communication \cite{ddgms}.

\subsection*{Beyond KZ scaling}

In section \ref{anyrate}, we were able to find the behaviour of the expectation value of the quenched operator when the coupling goes through its critical point at any finite time and any quench duration $\dt$. We did so numerically for both the fermionic TCP and scalar CCP quenches. Of course, when $m \dt$ is large and the time is near the critical time, we found that quenches obey Kibble-Zurek scaling. Further for the scalar CCPs, we found that same behaviour analytically with an expansion for large $\kappa=\sqrt{m\dt}$. As anticipated in eq.~\reef{t_over_tkz}, we found $\vev{\phi^2}_{ren} \sim (m/\dt)^{d-2/2} F(t/\tkz)$, where $F$ is a scaling function that only depends on $t/\tkz$. Apart from being a check of this KZ scaling formula, our calculations revealed a few other interesting aspects about the slow behaviour of quantum quenches: First, even though the natural dimensionless variable for the expansion in this regime is $1/\kappa$, it is straightforward to verify that corrections to the KZ scaling formula appear in powers of $1/\kappa^2=1/(m \dt)$ --- see Appendix \ref{subleading}.  In principle, this differs from previous holographic studies \cite{holo-kz} 
%\comment{more refs -- like in Sumit's email??} 
which found fractional powers of the quench rate. 
%\comment{Do you believe this?:} 
However, upon a closer examination, it appears that the corrections can be expressed in terms of simple integer powers of $1/\tkz$ in all of these different models. It would be interesting to develop a better analytic understanding of these corrections to the KZ scaling \reef{t_over_tkz}.
%
%This difference may point out some non-universal aspects of the slow quench regime. 

Second, and perhaps more surprising, is that $F(t/\tkz)$ is a good approximation to the full expectation value even beyond the Kibble-Zurek region, \ie $|t|/\tkz>1$. We found that for large $t/\tkz$, the scaling function $F$ takes a simple form which yields precisely the adiabatic response. This was a feature of the large $\k$ expansion and this agreement will fail for $|t|/\tkz>\k$. While $F(t/\tkz)$ will in general depend on the details of the theory and the quench protocol, it might have this universal characteristic of capturing both the KZ scaling and the adiabatic behaviour. It will be very interesting to extend these ideas and analytic findings first to the fermionic TCPs and then, try to formulate a similar expansion for general interacting CFTs, giving full evidence for universal scaling in slow quenches through a critical point.

\subsection*{`Not-quite-critical' slow quenches}

If we assume there is a CCP quench, we will find the corresponding KZ scaling for slow quenches near the instant where the protocol touches the critical point. Now one can ask what happens if the protocol does not `quite' touch the critical point but just goes close to it. To be more precise, we might be examine our scalar quenches with protocols of the form,
\beq
m^2(t) =  m^2 \tanh^{2}(t/\dt)+m^2\epsilon \,, \label{ugh}
\eeq
where $\epsilon \ll 1$. Rather than reaching the massless theory at $t=0$, this profile yields $m(t=0)= \sqrt{\epsilon}\,m$ (for $\epsilon>0$). Note that this is just the same CCP as in eq.~\reef{scalarquenchX}, but with a small shift in $m^2(t)$ by $m^2 \epsilon$. In particular, $\epsilon$ can be both positive or negative. Of course, if $\epsilon$ is some finite negative number (and $m\dt$ is large), then we should expect the system to become unstable with a negative mass-squared near $t=0$ (and everywhere if $\epsilon<-1$) and there will be a large number of particles created. However, if $|\epsilon|$ is small enough in magnitude then we might have some interesting behaviour. 

\begin{figure}[H]
        \centering
        \subfigure[$\epsilon<0$]{
                \includegraphics[scale=0.35]{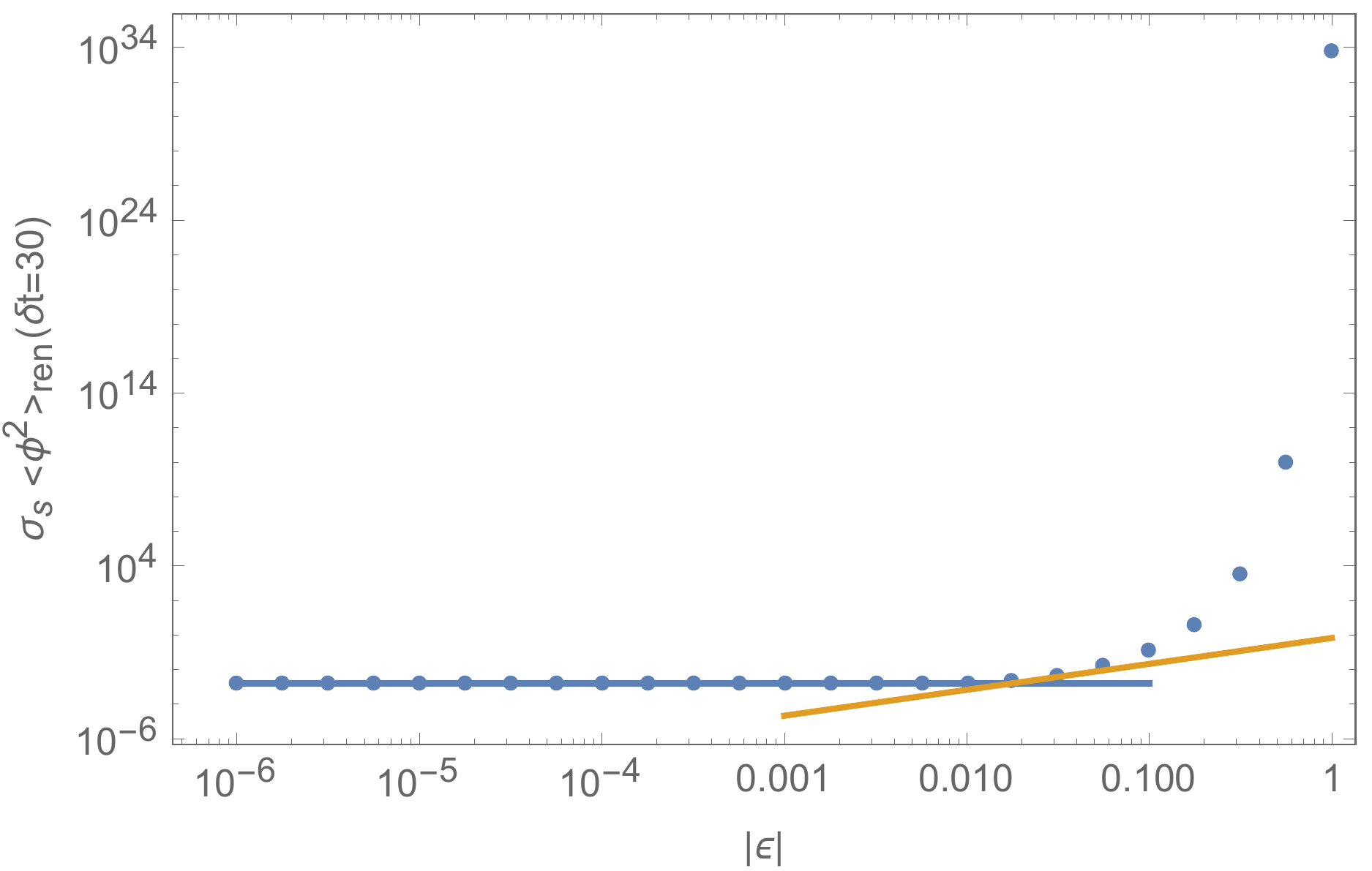} }
   		 \subfigure[$\epsilon>0$]{
                \includegraphics[scale=0.35]{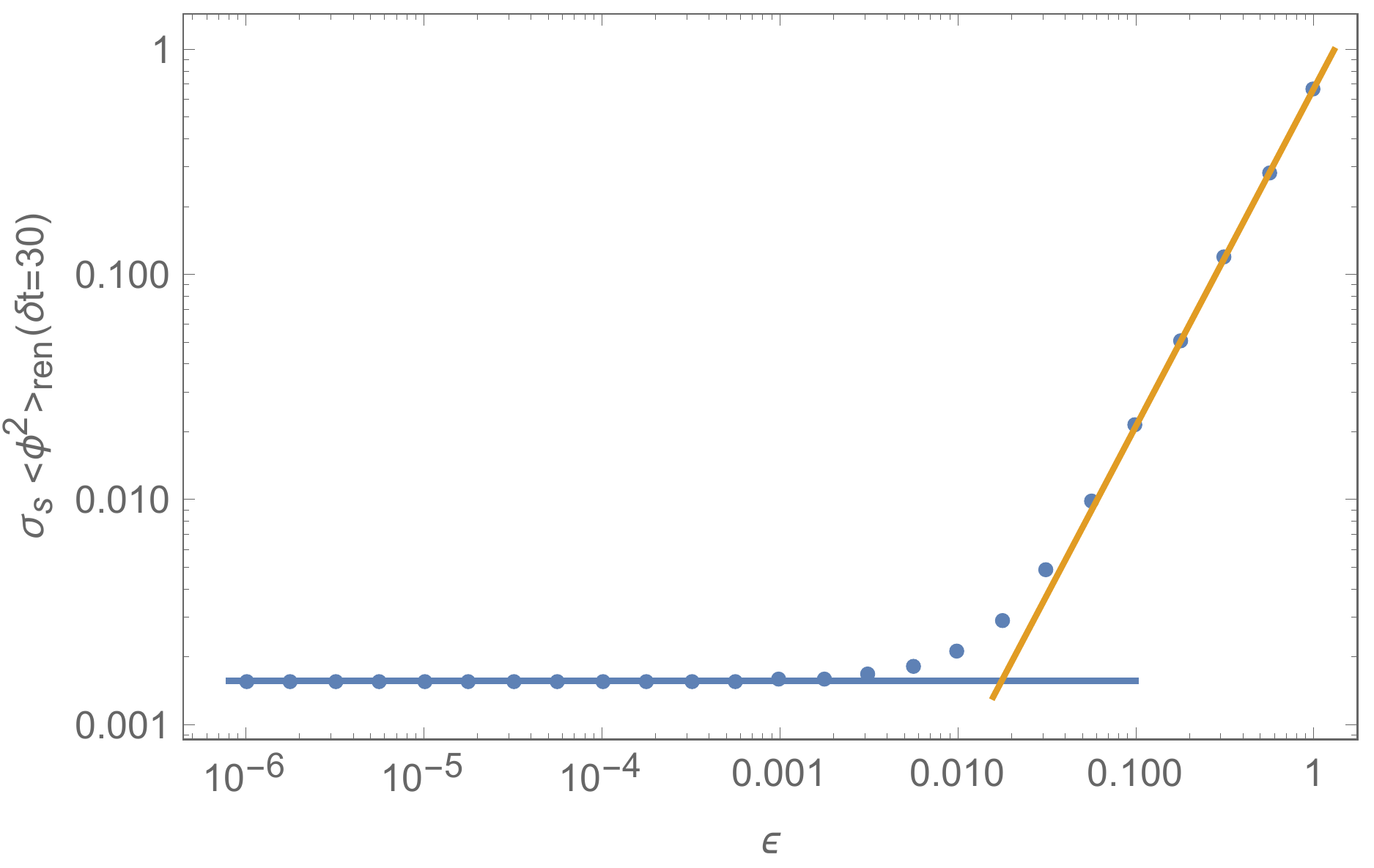} }
        \caption{Renormalizad expectation value of $\phi^2$ as a function of $\epsilon$ at $t=0$ for $d=5$. We are considering mass functions of the form $m^2(t)=m^2 \tanh^{2} (t/\dt)+m^2\epsilon$ and fixing $\dt=30$ and $m=1$. In panel (a) we show results for negative $\epsilon$, while in panel (b) we exhibit analogue results for positive $\epsilon$. In both cases, the blue solid line represents the KZ value at that $\dt$ and the yellow curve shows the adiabatic behaviour.
        } \label{fig_notquite}
\end{figure}

To study these `not-quite-critical' protocols, we fixed the time to zero and choose some large value for $m\dt$. Then we evaluate the expectation value as a function of  $\epsilon$. In figure \ref{fig_notquite}, we show the results for $d=5$ and $\dt=30$ (setting $m=1$). We see that for $|\epsilon| <1/\dt$, the expectation value is just the one expected from KZ scaling shown in eq. (\ref{an_d5}), independently of the sign of $\epsilon$. For positive $\epsilon$,  the expectation value takes its adiabatic value
when $|\epsilon| > 1/\dt$. In between, there is a smooth transition between the two regimes but apparently there is not any special scaling with $\epsilon$. On the contrary, when $\epsilon$ is large in magnitude compared to $1/\dt$ but negative, the expectation value rapidly diverges, pointing out the instability mentioned before.

The numerical results in figure \ref{fig_notquite} may not provide convincing evidence that the crossover between the two regimes occurs at $|\epsilon| \simeq 1/(m\dt)$ --- where the factor of $1/m$ to make the result appropriately dimensionless. Hence let  us estimate where the transition takes place by finding the intersection of the curves describing the KZ scaling and the adiabatic expectation value. From eq.~(\ref{an_d5}), we estimate $\vev{\phi^2}_{ren}\approx 0.25\,(m/\dt)^{3/2}$ at $t=0$. For the mass profile in eq. (\ref{ugh}) with $\epsilon > 0$, the adiabatic expectation value at $t=0$ becomes $\frac{2}{3} m^{3} \epsilon^{3/2}$. %\comment{I guess we are assuming $\epsilon>0$??} 
The intersection between the two gives that the transition should be at order $\epsilon \simeq 1/(m\dt)$, which was our estimate from figure (\ref{fig_notquite}). That is, we have found that the KZ scaling survives as long as $m(t=0)=\sqrt{\epsilon}\,m\le  1/\tkz = \sqrt{m/\dt}$.

A more precise explanation of this behaviour can be found by examining the condition for the breakdown of adiabaticity in eq.~\reef{1-1}: $\frac{1}{m(t)^2}\frac{d m(t)}{dt} \simeq 1$. For the following analysis, we limit our attention to $\epsilon\ge0$ for simplicity. 
Now let us assume that adiabaticity breaks at some $t = \tkz(\epsilon)$ with $\tkz(\epsilon) \ll \dt$. The latter implies that in the relevant regime, the mass profile (\ref{ugh}) can be simplied to: 
\ben 
m^2(t) =  m^2 \,\left[(t/\dt)^2+\epsilon\right]\,.
\labell{ugh2}
\een 
Then, with a change of variables, 
the adiabaticity condition may be written as
\ben
\frac{y}{(m  \dt\,\epsilon)^{2}} \simeq (1+y)^3 \qquad{\rm where}\ \ \ y = \frac{t^2}{\epsilon (\dt)^2}\,.
\een
It is clear that this condition cannot be met if $\epsilon$ becomes too large. The critical value $\epsilon_c$ (such that adiabaticity breaks for $\epsilon < \epsilon_c$) is obtained when the line $(m \epsilon \dt )^{-2} y$ becomes tangent to the curve $(1+y)^3$. This is easily determined to be
\ben
\epsilon_c = \frac{2}{3\sqrt{3}\,m \dt} \labell{exotic}
\een
The condition $\epsilon < \epsilon_c$ is clearly the same as the condition given above upto a numerical factor.  

Now it is straightforward to show that for $0<\epsilon \le \epsilon_c$, adiabaticity breaks at some time $\tkz(\epsilon)<\tkz(0)=\sqrt{\dt/m}$. 
Further, at the critical value $\epsilon = \epsilon_c$, the Kibble-Zurek time is given by $\tkz(\epsilon_c) = \sqrt{\epsilon_c/2}\,\dt$. Hence using eq.~\reef{exotic}, we have
\beq
 \frac{\sqrt{\dt/m}}{3\sqrt{3}}\,\le\tkz(\epsilon)\le\sqrt{\dt/m}\,. \labell{lake}
\eeq
Hence for any value of $\epsilon$, we have $\tkz \sim \dt/\sqrt{m\dt}\ll \dt$ since we are in the slow quench regime. This justifies the initial assumption made above eq.~\reef{ugh2}. If all of the other parameters are held fixed, the mass value at $\tkz(\epsilon)$ also decreases as $\epsilon$ increases. One can express the general result as
\beq
m(\tkz(\epsilon))=\frac1{\tkz(0)}\,\left(\frac{\tkz(\epsilon)}{\tkz(0)}\right)^{1/3}\,.
\eeq

Given that properties of the non-adiabatic regime are changing quick significantly as $\epsilon$ varies, it may seem surprising that na\"ive KZ scaling appears to fit so well in figure \ref{fig_notquite}. However, this is largely because $\epsilon$ is negligibly small over most of the range where the blue line fits the numerical results. Note that for positive $\epsilon$ in panel (b), we can see that the numerical result already shows an appreciable difference at $\epsilon\sim0.006$, while $\epsilon_c\simeq0.013$ for the parameters used in the figure.

\section*{Acknowledgements} S.R.D. would like to thank D. Das, H. Liu, G. Mandal, K. Sengupta and participants of the workshop ``The Non-equilibrium quantum frontier" at Princeton University for discussions. 
The work of SRD is partially supported by the National Science Foundation grants NSF-PHY-1214341 and NSF-PHY-1521045.
Research at Perimeter Institute is supported by the
Government of Canada through the Department of Innovation, Science and Economic Development and by the Province of Ontario
through the Ministry of Research \& Innovation. RCM and DAG are also supported
by an NSERC Discovery grant. RCM is also supported by research funding
from the Canadian Institute for Advanced Research and from the Simons Foundation through ``It from Qubit" Collaboration.
DAG also thanks the Kavli Institute for Theoretical Physics, where he was a Visiting Graduate Fellow, for hospitality and support during the first stages of this project. Research at KITP is supported, in part, by the National Science Foundation under Grant No. NSF PHY11-25915.

\newpage

\appendix

\section{Subleading contributions to KZ scaling}
\label{subleading}
In the main text, we discussed the emergence of KZ scaling in the expectation values of quenched operators after slow quenches that go through a critical point. In particular, for protocols where the mass approaches the critical point  linearly, it was possible to define a KZ time, $\tkz = \sqrt{\dt/m}$, as in eq.~\reef{ttkz}. Then, the leading behaviour of corresponding expectation values is controlled by $\tkz$ as in eq.~\reef{t_over_tkz}. The aim of this Appendix is to analyze the subleading corrections to this scaling behaviour 
\begin{eqnarray}
\vev{\calo_\Delta}_{ren} = \tkz^{-\Delta} \left[ F(t/\tkz) + \frac1{(m\dt)^\alpha}\,F_2(t/\tkz) +\cdots \right] \,. \label{appendix}
\end{eqnarray}
In the above expression, we have anticipated that the corrections to the scaling function are controlled by the combination $m\dt$, which is the only dimensionless parameter which characterizes the quench. The key question which we will address here is determining the power $\alpha$ for our free field quenches.  Some previous studies \cite{holo-kz} found that in a variety of holographic models, these corrections come in fractional powers of the quench duration $\dt$.

In section \ref{scalingfns}, we constructed the scaling function $F(t/\tkz)$ for the scalar CCP quenches. Our approach there was to expand the full solution for large $\kappa = \sqrt{m \dt}$ while holding $t/\tkz$ fixed. A natural first guess might then be that the first subleading correction to $F(t/\tkz)$ should be of order $1/\kappa =1/\sqrt{m \dt}$. However, this turns out not to be the case. In fact,  the first correction comes at order $1/\kappa^{2}=1/(m\dt)$ for both the scalar CCP and fermionic TCP quenches. In the scalar field case, this result can be determined by simply examining the different factors appearing in eq.~(\ref{golly}) and looking at their expansion for large $\kappa$. In doing so, we realize that all the different corrections to the leading scaling behaviour appear at order $1/\kappa^{2}$. In the following, we will extract that this leading correction numerically for both the fermionic TCP and scalar CCP quenches. 

\begin{figure}[H]
        \centering
        \subfigure[Fermionic TCP]{
                \includegraphics[scale=0.38]{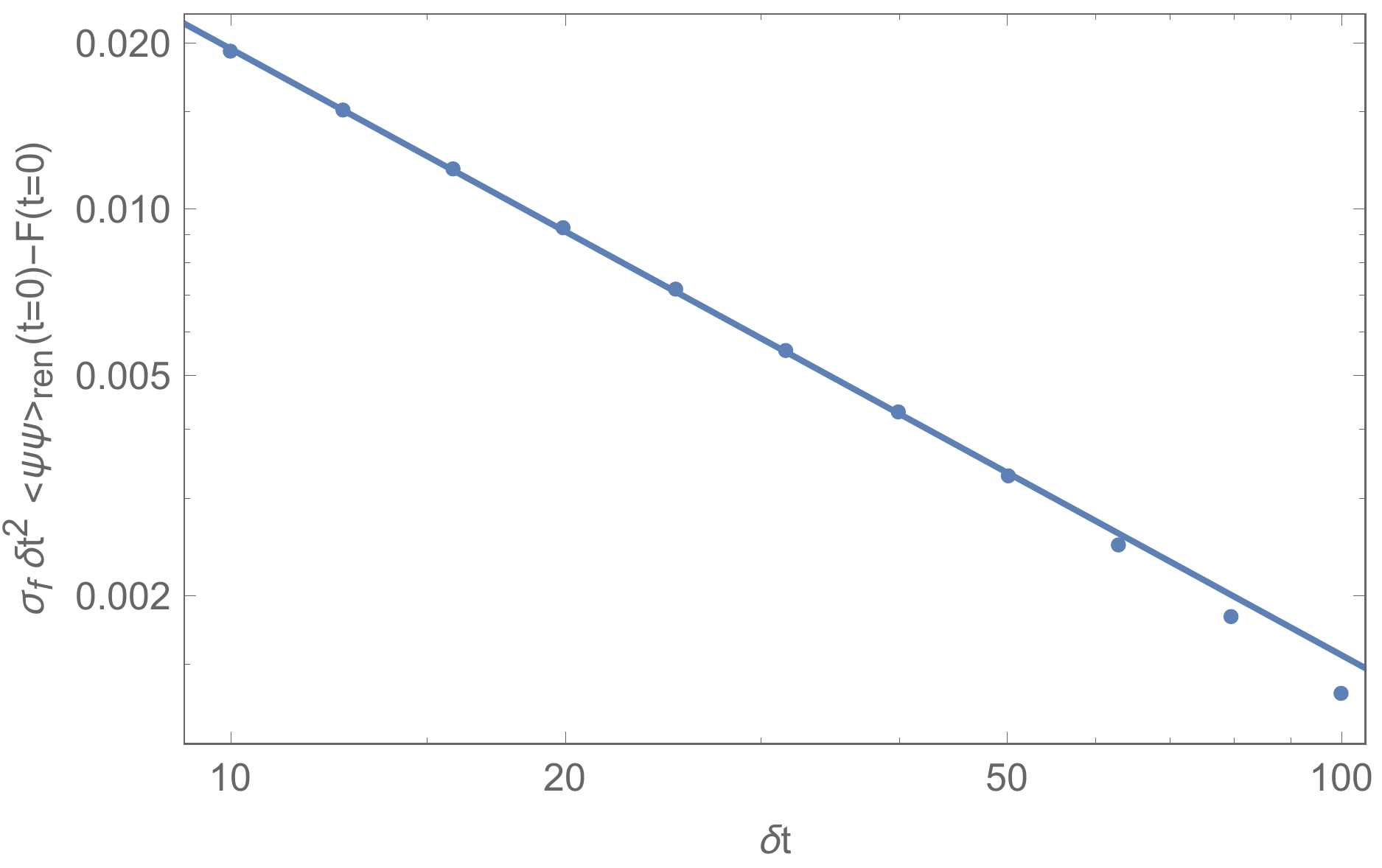} }
   		 \subfigure[Scalar CCP]{
                \includegraphics[scale=0.38]{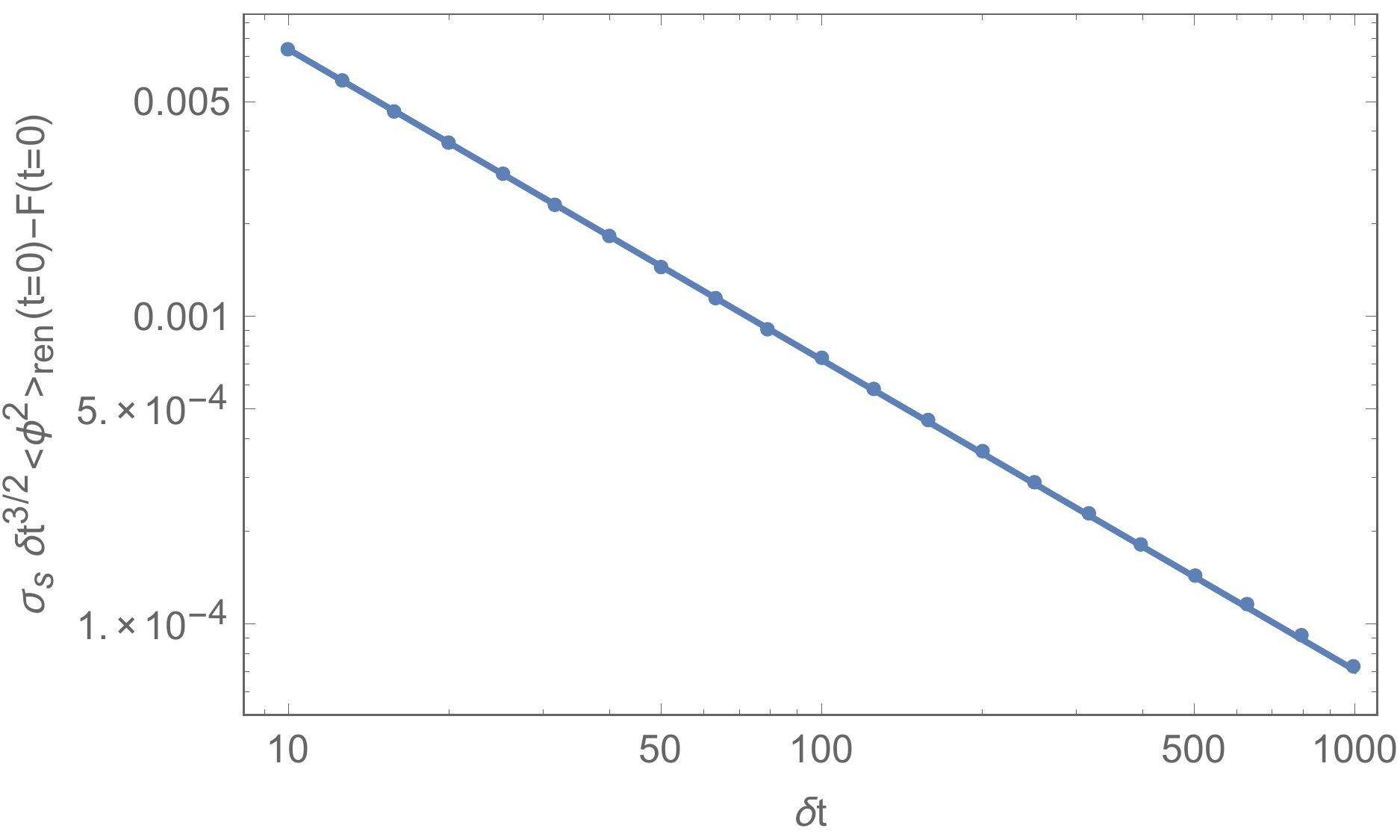} }
        \caption{Subleading contributions to the expectation value of mass operators in fermionic TCP and scalar CCP quenches at $t/\tkz=0$, with $d=5$ and $m=1$. In each case, we are multiplying the full expectation value by the expected KZ scaling and then subtracting $F(0)$. The remainder are the subleading contributions, that we fit as a function of $\dt$: $y=a \dt^{\alpha}$. The result of the fit (in solid blue) gives $\alpha=-1.09684$ in the fermionic case and $\alpha= -1.00941$ for scalar case, both supporting the idea that the first corrections to KZ scaling in free field theory come at order $1/\dt$.
        } \label{fig_appendix}
\end{figure}

Next, we consider the fermionic TCP quenches in $d=5$. We numerically compute the expectation value at a fixed $t/\tkz$ and large $m\dt$ and perform a fit with the expected KZ scaling, \eg as in figure \ref{fig_kz_zero_ferm}. The overall coefficient in this fit yields our numerical estimate of the scaling function $F(t/\tkz)$ at the given value of $t/\tkz$. \beq
\frac1{(m\dt)^\alpha}\,F_2(t/\tkz) = \frac{\dt^2}{m^2}\, \sigma_f \vev{\bar{\psi}\psi}_{ren} (t/\tkz) - F(t/\tkz) \labell{wack}
\eeq 
as a function of $m\dt$ to identify the scaling power $\alpha$. Note that the factor $(\dt^2/m)^2$ on the right-hand side simply cancels the expected KZ scaling \reef{KZscalingTCP} (for $d=5$) of the expectation value. The numerical fit of the results shown in figure \ref{fig_appendix} shows that this first subleading contribution scales as $1/\dt$ at $t/\tkz=0$. One might worry that the $F_2(t/\tkz)$ vanishes at $t=0$. However, performing the analogous calculations at a different finite values of $t/\tkz$ yields the subleading scaling, \ie eq.~\reef{wack} scales as $1/\dt$. Hence we conclude that the corrections to the KZ scaling in these free field quenches take the form given in eq. (\ref{appendix}) with $\alpha=1$.

Of course, as described above for the scalar CCP quenches, we can reveal this $1/\dt$ scaling analytically from our construction of $F(t/\tkz)$ in section \ref{scalingfns}. However, we can also use the scaling function \reef{analytic} in an analogous numerical calculation as described above for the fermionic TCP quenches.  In particular, in $d=5$, we numerically evaluated $(\dt/m)^{3/2} \sigma_s \vev{\phi^2}_{ren} (t/\tkz) - F(t/\tkz)$  for fixed $t/\tkz$ as a function of $m\dt$. The results are shown in figure (\ref{fig_appendix}) for $t/\tkz=0$. Numerically fitting these results shows that the subleading correction scales as $1/\dt$, as expected. Hence we again find that the corrections to the KZ scaling take the form in eq. (\ref{appendix}) with $\alpha=1$.

This clearly shows that the subleading corrections for our free field quenches come with integer powers of $\dt$. Of course, this contrasts with the previous holographic studies \cite{holo-kz}, 
%\comment{more refs -- like in Sumit's email??} 
which found fractional powers of the quench duration.

\end{document}